\pdfoutput=1
\documentclass[12pt,a4paper]{article}
\usepackage{ifthen} % for conditional statements
\newboolean{pdflatex}
\setboolean{pdflatex}{true} % False for eps figures 

\newboolean{articletitles}
\setboolean{articletitles}{true} % False removes titles in references

\newboolean{uprightparticles}
\setboolean{uprightparticles}{false} %True for upright particle symbols

\newboolean{inbibliography}
\setboolean{inbibliography}{false} %True once you enter the bibliography

\usepackage{microtype}
\usepackage{lineno}  % for line numbering during review
\usepackage{xspace} % To avoid problems with missing or double spaces after
\usepackage{graphicx}  % to include figures (can also use other packages)
\usepackage{color}
\usepackage{colortbl}
\graphicspath{{./figs/}} % Make Latex search fig subdir for figures

\usepackage{amsmath} % Adds a large collection of math symbols
\usepackage{amssymb}
\usepackage{amsfonts}
\usepackage{upgreek} % Adds in support for greek letters in roman typeset

\newcommand*\patchAmsMathEnvironmentForLineno[1]{%
\expandafter\let\csname old#1\expandafter\endcsname\csname #1\endcsname
\expandafter\let\csname oldend#1\expandafter\endcsname\csname
end#1\endcsname
 \renewenvironment{#1}%
   {\linenomath\csname old#1\endcsname}%
   {\csname oldend#1\endcsname\endlinenomath}%
}
\newcommand*\patchBothAmsMathEnvironmentsForLineno[1]{%
  \patchAmsMathEnvironmentForLineno{#1}%
  \patchAmsMathEnvironmentForLineno{#1*}%
}
\AtBeginDocument{%
\patchBothAmsMathEnvironmentsForLineno{equation}%
\patchBothAmsMathEnvironmentsForLineno{align}%
\patchBothAmsMathEnvironmentsForLineno{flalign}%
\patchBothAmsMathEnvironmentsForLineno{alignat}%
\patchBothAmsMathEnvironmentsForLineno{gather}%
\patchBothAmsMathEnvironmentsForLineno{multline}%
}

\usepackage{hyperref}    % Hyperlinks in references
\usepackage[all]{hypcap} % Internal hyperlinks to floats.

\def\lhcb {\mbox{LHCb}\xspace}

\ifthenelse{\boolean{uprightparticles}}%
{
 
 \def\Pgamma      {\ensuremath{\upgamma}\xspace}

 \def\Pmu         {\ensuremath{\upmu}\xspace}

 \def\Ppi         {\ensuremath{\uppi}\xspace}

 \def\Ppsi        {\ensuremath{\uppsi}\xspace}

 \def\PDelta      {\ensuremath{\Delta}\xspace}                 
 \def\PXi      {\ensuremath{\Xi}\xspace}                 
 \def\PLambda      {\ensuremath{\Lambda}\xspace}                 
 \def\PSigma      {\ensuremath{\Sigma}\xspace}                 
 \def\POmega      {\ensuremath{\Omega}\xspace}                 
 \def\PUpsilon      {\ensuremath{\Upsilon}\xspace}                 
 
 \def\PB      {\ensuremath{\mathrm{B}}\xspace}                 
                  
 \def\PD      {\ensuremath{\mathrm{D}}\xspace}

 \def\PJ      {\ensuremath{\mathrm{J}}\xspace}                 
 \def\PK      {\ensuremath{\mathrm{K}}\xspace}

 \def\Pb      {\ensuremath{\mathrm{b}}\xspace}                 
 \def\Pc      {\ensuremath{\mathrm{c}}\xspace}                 
 \def\Pd      {\ensuremath{\mathrm{d}}\xspace}

 \def\Pi      {\ensuremath{\mathrm{i}}\xspace}

 \def\Pp      {\ensuremath{\mathrm{p}}\xspace}

 \def\Ps      {\ensuremath{\mathrm{s}}\xspace}                 
                  
 \def\Pu      {\ensuremath{\mathrm{u}}\xspace}

}
{
 
 \def\Pgamma      {\ensuremath{\gamma}\xspace}

 \def\Pmu         {\ensuremath{\mu}\xspace}

 \def\Ppi         {\ensuremath{\pi}\xspace}

 \def\Ppsi        {\ensuremath{\psi}\xspace}                 
                  
 \mathchardef\PDelta="7101
 \mathchardef\PXi="7104
 \mathchardef\PLambda="7103
 \mathchardef\PSigma="7106
 \mathchardef\POmega="710A
 \mathchardef\PUpsilon="7107
                  
 \def\PB      {\ensuremath{B}\xspace}                 
                  
 \def\PD      {\ensuremath{D}\xspace}

 \def\PJ      {\ensuremath{J}\xspace}                 
 \def\PK      {\ensuremath{K}\xspace}

 \def\Pb      {\ensuremath{b}\xspace}                 
 \def\Pc      {\ensuremath{c}\xspace}                 
 \def\Pd      {\ensuremath{d}\xspace}

 \def\Pi      {\ensuremath{i}\xspace}

 \def\Pp      {\ensuremath{p}\xspace}

 \def\Ps      {\ensuremath{s}\xspace}                 
                  
 \def\Pu      {\ensuremath{u}\xspace}

}

\def\mumu       {\ensuremath{\Pmu^+\Pmu^-}\xspace}

\def\g      {\ensuremath{\Pgamma}\xspace}

\def\uquark    {\ensuremath{\Pu}\xspace}
\def\uquarkbar {\ensuremath{\overline \uquark}\xspace}

\def\dquark    {\ensuremath{\Pd}\xspace}

\def\squark    {\ensuremath{\Ps}\xspace}

\def\cquark    {\ensuremath{\Pc}\xspace}

\def\bquark    {\ensuremath{\Pb}\xspace}

\def\pion  {\ensuremath{\Ppi}\xspace}
\def\piz   {\ensuremath{\pion^0}\xspace}

\def\pip   {\ensuremath{\pion^+}\xspace}
\def\pim   {\ensuremath{\pion^-}\xspace}

\def\kaon  {\ensuremath{\PK}\xspace}
  \def\Kbar  {\kern 0.2em\overline{\kern -0.2em \PK}{}\xspace}

\def\Kp    {\ensuremath{\kaon^+}\xspace}
\def\Km    {\ensuremath{\kaon^-}\xspace}

\def\Kstarz  {\ensuremath{\kaon^{*0}}\xspace}

  \def\Dbar    {\kern 0.2em\overline{\kern -0.2em \PD}{}\xspace}
\def\D       {\ensuremath{\PD}\xspace}

\def\Dz      {\ensuremath{\D^0}\xspace}
\def\Dzb     {\ensuremath{\Dbar^0}\xspace}
\def\Dp      {\ensuremath{\D^+}\xspace}

\def\Dstarz  {\ensuremath{\D^{*0}}\xspace}

\def\Dstarp  {\ensuremath{\D^{*+}}\xspace}

\def\Ds      {\ensuremath{\D^+_\squark}\xspace}

\def\B       {\ensuremath{\PB}\xspace}
\def\Bbar    {\ensuremath{\kern 0.18em\overline{\kern -0.18em \PB}{}}\xspace}

\def\Bz      {\ensuremath{\B^0}\xspace}
\def\Bzb     {\ensuremath{\Bbar^0}\xspace}
\def\Bu      {\ensuremath{\B^+}\xspace}

\def\Bp      {\ensuremath{\Bu}\xspace}

\def\Bsb     {\ensuremath{\Bbar^0_\squark}\xspace}

\def\jpsi     {\ensuremath{{\PJ\mskip -3mu/\mskip -2mu\Ppsi\mskip 2mu}}\xspace}

  \def\Y#1S{\ensuremath{\PUpsilon{(#1S)}}\xspace}% no space before {...}!

\def\proton      {\ensuremath{\Pp}\xspace}

\def\Xires {\ensuremath{\PXi}\xspace}

\def\Lz {\ensuremath{\PLambda}\xspace}
\def\Lbar {\ensuremath{\kern 0.1em\overline{\kern -0.1em\PLambda}}\xspace}

\def\Sigmares {\ensuremath{\PSigma}\xspace}

\def\Omegares {\ensuremath{\POmega^-}\xspace}

\def\Lb      {\ensuremath{\Lz^0_\bquark}\xspace}

\def\Lc      {\ensuremath{\Lz^+_\cquark}\xspace}

\def\BF         {{\ensuremath{\cal B}\xspace}}

\def\BR         {\BF}
\def\to                 {\ensuremath{\rightarrow}\xspace}

\def\CP                {\ensuremath{C\!P}\xspace}

\def\AT#1     {\ensuremath{A_{\mathrm{T}}^{#1}}\xspace}           % 2

\def\C#1      {\ensuremath{\mathcal{C}_{#1}}\xspace}                       % 9
\def\Cp#1     {\ensuremath{\mathcal{C}_{#1}^{'}}\xspace}                    % 7
\def\Ceff#1   {\ensuremath{\mathcal{C}_{#1}^{\mathrm{(eff)}}}\xspace}        % 9  
\def\Cpeff#1  {\ensuremath{\mathcal{C}_{#1}^{'\mathrm{(eff)}}}\xspace}       % 7
\def\Ope#1    {\ensuremath{\mathcal{O}_{#1}}\xspace}                       % 2
\def\Opep#1   {\ensuremath{\mathcal{O}_{#1}^{'}}\xspace}                    % 7

\newcommand{\tev}{\ifthenelse{\boolean{inbibliography}}{\ensuremath{~T\kern -0.05em eV}\xspace}{\ensuremath{\mathrm{\,Te\kern -0.1em V}}\xspace}}
\newcommand{\gev}{\ensuremath{\mathrm{\,Ge\kern -0.1em V}}\xspace}
\newcommand{\mev}{\ensuremath{\mathrm{\,Me\kern -0.1em V}}\xspace}
\newcommand{\kev}{\ensuremath{\mathrm{\,ke\kern -0.1em V}}\xspace}
\newcommand{\ev}{\ensuremath{\mathrm{\,e\kern -0.1em V}}\xspace}
\newcommand{\gevc}{\ensuremath{{\mathrm{\,Ge\kern -0.1em V\!/}c}}\xspace}
\newcommand{\mevc}{\ensuremath{{\mathrm{\,Me\kern -0.1em V\!/}c}}\xspace}
\newcommand{\gevcc}{\ensuremath{{\mathrm{\,Ge\kern -0.1em V\!/}c^2}}\xspace}
\newcommand{\gevgevcccc}{\ensuremath{{\mathrm{\,Ge\kern -0.1em V^2\!/}c^4}}\xspace}
\newcommand{\mevcc}{\ensuremath{{\mathrm{\,Me\kern -0.1em V\!/}c^2}}\xspace}

\def\mum  {\ensuremath{\,\upmu\rm m}\xspace}

\def\invfb   {\ensuremath{\mbox{\,fb}^{-1}}\xspace}

\def\ps   {\ensuremath{{\rm \,ps}}\xspace}

\newcommand{\chisq}{\ensuremath{\chi^2}\xspace}

\newcommand{\chisqip}{\ensuremath{\chi^2_{\rm IP}}\xspace}

\def\gsim{{~\raise.15em\hbox{$>$}\kern-.85em
          \lower.35em\hbox{$\sim$}~}\xspace}
\def\lsim{{~\raise.15em\hbox{$<$}\kern-.85em
          \lower.35em\hbox{$\sim$}~}\xspace}

\def\sPlot{\mbox{\em sPlot}}

\def\pt         {\mbox{$p_{\rm T}$}\xspace}

\def\dllkpi     {\ensuremath{\mathrm{DLL}_{\kaon\pion}}\xspace}
\def\dllppi     {\ensuremath{\mathrm{DLL}_{\proton\pion}}\xspace}
\def\dllpk      {\ensuremath{\mathrm{DLL}_{\proton\kaon}}\xspace}

\def\evtgen     {\mbox{\textsc{EvtGen}}\xspace}

\def\geant      {\mbox{\textsc{Geant4}}\xspace}

\def\pythia     {\mbox{\textsc{Pythia}}\xspace}

\def\tell1  {TELL1\xspace}
\def\ukl1   {UKL1\xspace}

\newcommand{\ie}{\mbox{\itshape i.e.}\xspace}

\newcommand{\xibn}{\ensuremath{\Xires_{\bquark}^0}\xspace}
\newcommand{\lblcpi}{\ensuremath{\Lb\to\Lc\pim}\xspace}
\newcommand{\lblck}{\ensuremath{\Lb\to\Lc\Km}\xspace}
\newcommand{\lblch}{\ensuremath{\Lb\to\Lc h^-}\xspace}
\newcommand{\lbdppi}{\ensuremath{\Lb\to \Dz \proton \pim}\xspace}
\newcommand{\lbdpk}{\ensuremath{\Lb\to \Dz \proton \Km}\xspace}
\newcommand{\lbdph}{\ensuremath{\Lb\to \Dz \proton h^-}\xspace}
\newcommand{\xibdpk}{\ensuremath{\xibn\to \Dz \proton \Km}\xspace}
\newcommand{\xiblck}{\ensuremath{\xibn\to \Lc \Km}\xspace}
\newcommand{\lcpi}{\ensuremath{\Lc\pim}\xspace}
\newcommand{\lck}{\ensuremath{\Lc\Km}\xspace}
\newcommand{\lch}{\ensuremath{\Lc h^-}\xspace}
\newcommand{\dppi}{\ensuremath{\Dz \proton \pim}\xspace}
\newcommand{\dpk}{\ensuremath{\Dz \proton \Km}\xspace}
\newcommand{\dph}{\ensuremath{\Dz \proton h^-}\xspace}
\newcommand{\dbppi}{\ensuremath{\Dzb \proton \pim}\xspace}
\newcommand{\dbpk}{\ensuremath{\Dzb \proton \Km}\xspace}
\newcommand{\dbph}{\ensuremath{\Dzb \proton h^-}\xspace}

\newcommand{\lbdbpk}{\ensuremath{\Lb\to \Dzb \proton \Km}\xspace}

\newcommand{\lcpkpi}{\ensuremath{\Lc\to \proton \Km\pip}\xspace}
\newcommand{\pkpi}{\ensuremath{\proton \Km\pip}\xspace}
\newcommand{\dnkpi}{\ensuremath{\Dz\to \Km\pip}\xspace}

\usepackage{cite} % Allows for ranges in citations
\usepackage{mciteplus}

\usepackage{longtable} % only for template; not usually to be used in PAPERs

\begin{document}

\renewcommand{\thefootnote}{\fnsymbol{footnote}}
\setcounter{footnote}{1}

% $Id: title-LHCb-PAPER.tex 35201 2013-05-10 14:33:25Z roldeman $
% ===============================================================================
% Purpose: LHCb-PAPER journal paper title page template
% Author: 
% Created on: 2010-09-25
% ===============================================================================

%%%%%%%%%%%%%%%%%%%%%%%%%
%%%%%  TITLE PAGE  %%%%%%
%%%%%%%%%%%%%%%%%%%%%%%%%
\begin{titlepage}
\pagenumbering{roman}

% Header ---------------------------------------------------
\vspace*{-1.5cm}
\centerline{\large EUROPEAN ORGANIZATION FOR NUCLEAR RESEARCH (CERN)}
\vspace*{1.5cm}
\hspace*{-0.5cm}
\begin{tabular*}{\linewidth}{lc@{\extracolsep{\fill}}r}
\vspace*{-2.7cm}\mbox{\!\!\!\includegraphics[width=.14\textwidth]{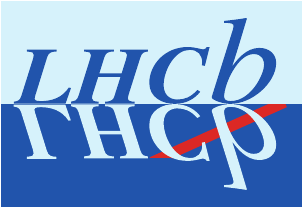}} & &
\\
 & & CERN-PH-EP-2013-207 \\  % ID 
 & & LHCb-PAPER-2013-056 \\  % ID 
 & & 19 November 2013 \\ % Date - Can also hardwire e.g.: 23 March 2010
 & & \\
% not in paper \hline
\end{tabular*}

\vspace*{4.0cm}

% Title --------------------------------------------------
{\bf\boldmath\huge
\begin{center}
  Studies of beauty baryon decays to \dph and \lch final states
\end{center}
}

\vspace*{2.0cm}

% Authors -------------------------------------------------
\begin{center}
The LHCb collaboration\footnote{Authors are listed on the following pages.}
\end{center}

\vspace{\fill}

% Abstract -----------------------------------------------
\begin{abstract}
  \noindent
  
  Decays of beauty baryons to the \dph and \lch final states 
  (where $h$ indicates a pion or a kaon) are studied using a data sample of $\proton\proton$
  collisions, corresponding to an integrated 
  luminosity of 1.0\invfb, collected by the LHCb detector. 
  The Cabibbo-suppressed decays \lbdpk and \lblck
  are observed and their branching fractions are measured with respect to 
  the decays \lbdppi and \lblcpi. In addition, the first observation is reported 
  of the decay of the neutral beauty-strange baryon \xibn to
  the \dpk final state, and a measurement of the \xibn mass is performed. 
  Evidence of the \xiblck decay is also reported. 
\end{abstract}

\vspace*{2.0cm}

\begin{center}
  Submitted to Phys.~Rev.~D
\end{center}

\vspace{\fill}

{\footnotesize 
\centerline{\copyright~CERN on behalf of the \lhcb collaboration, license \href{http://creativecommons.org/licenses/by/3.0/}{CC-BY-3.0}.}}
\vspace*{2mm}

\end{titlepage}

%%%%%%%%%%%%%%%%%%%%%%%%%%%%%%%%
%%%%%  EOD OF TITLE PAGE  %%%%%%
%%%%%%%%%%%%%%%%%%%%%%%%%%%%%%%%

%  empty page follows the title page ----
\newpage
\setcounter{page}{2}
\mbox{~}
\newpage

%%%%%%%%%%%%%%%%%%%%%%%%%%%%%%%%%%%%%%%%%%
\centerline{\large\bf LHCb collaboration}
\begin{flushleft}
\small
R.~Aaij$^{40}$, 
B.~Adeva$^{36}$, 
M.~Adinolfi$^{45}$, 
C.~Adrover$^{6}$, 
A.~Affolder$^{51}$, 
Z.~Ajaltouni$^{5}$, 
J.~Albrecht$^{9}$, 
F.~Alessio$^{37}$, 
M.~Alexander$^{50}$, 
S.~Ali$^{40}$
% ...

{\footnotesize \it
% ...
$ ^{5}$Clermont Universit\'{e}, Universit\'{e} Blaise Pascal, CNRS/IN2P3, LPC, Clermont-Ferrand, France\\
$ ^{6}$CPPM, Aix-Marseille Universit\'{e}, CNRS/IN2P3, Marseille, France\\
% ...
$ ^{9}$Fakult\"{a}t Physik, Technische Universit\"{a}t Dortmund, Dortmund, Germany\\
% ...
$ ^{36}$Universidad de Santiago de Compostela, Santiago de Compostela, Spain\\
$ ^{37}$European Organization for Nuclear Research (CERN), Geneva, Switzerland\\
% ...
$ ^{40}$Nikhef National Institute for Subatomic Physics, Amsterdam, The Netherlands\\
% ...
$ ^{45}$H.H. Wills Physics Laboratory, University of Bristol, Bristol, United Kingdom\\
% ...
$ ^{50}$School of Physics and Astronomy, University of Glasgow, Glasgow, United Kingdom\\
$ ^{51}$Oliver Lodge Laboratory, University of Liverpool, Liverpool, United Kingdom\\
% ...
}
\end{flushleft}
%%%%%%%%%%%%%%%%%%%%%%%%%%%%%%%%%%%%%%%%%%

\cleardoublepage

\renewcommand{\thefootnote}{\arabic{footnote}}
\setcounter{footnote}{0}

\pagestyle{plain} % restore page numbers for the main text
\setcounter{page}{1}
\pagenumbering{arabic}

%\linenumbers

\section{Introduction}
\label{sec:Introduction}

Although there has been great progress in studies of beauty mesons, both at the 
\B factories and hadron machines, the beauty baryon sector remains largely unexplored. 
The quark model predicts seven ground-state
($J^P = \frac{1}{2}^{+}$) baryons involving a \bquark quark and two light
(\uquark, \dquark, or \squark) quarks~\cite{PDG2012}. These are the \Lb isospin singlet, the
$\Sigmares_{\bquark}$ triplet, the $\Xires_{\bquark}$ strange doublet, and the 
doubly strange state $\Omegares_{\bquark}$. Among these states, 
the $\Sigmares_{\bquark}^0$ baryon has not been observed yet, 
while for the others the quantum numbers have not been experimentally established, 
very few decay modes have been measured, and fundamental properties such as masses and
lifetimes are in general poorly known. Moreover, the 
$\Sigmares_{\bquark}^{\pm}$ and $\Xires_{\bquark}^0$ baryons have been observed by a single 
experiment~\cite{:2007rw, Aaltonen:2011wd}. It is therefore of great interest to 
study \bquark baryons, and to determine their properties. 

The decays of \bquark baryons can be used to study \CP violation and rare processes. 
In particular, the decay $\Lb\to\Dz\Lz$ has been proposed to measure the
Cabibbo-Kobayashi-Maskawa (CKM) unitarity triangle angle $\gamma$~\cite{Dunietz:1992ti,:1998upb,Giri:2001ju}
following an approach analogous to that for $\Bz\to\D\Kstarz$ decays~\cite{Dunietz:1991yd}. 
A possible extension to the analysis of the $\Dz\Lz$ final state is
to use the $\lbdpk$ decay, with the
$\proton\Km$ pair originating from the \Lb decay vertex. 
Such an approach can avoid limitations due to the lower reconstruction
efficiency of the $\Lz$ decay. In addition, if the full
phase space of the three-body decay is used, the
sensitivity to $\gamma$ may be enhanced, in a similar manner to the Dalitz plot
analysis of $\Bz \to \D \Kp\pim$ decays, which offers certain advantages over 
the quasi-two-body $\Bz \to \D \Kstarz$ analysis~\cite{Gershon:2008pe,Gershon:2009qc}.

This paper reports the results of a study of beauty baryon decays into  
\dppi, \dpk, \lcpi, and \lck final states.\footnote{The inclusion of charge-conjugate 
processes is implied.}
A data sample corresponding to an integrated luminosity of 1.0\invfb is used, collected by the 
LHCb detector~\cite{Alves:2008zz} in $\proton\proton$ collisions with 
centre-of-mass energy of 7\tev. Six measurements are performed in this analysis, listed below.

The decay mode \lbdppi is the Cabibbo-favoured partner of \lbdpk with the same topology 
and higher rate. We measure its rate using the mode \lblcpi for normalisation. 
To avoid dependence on the poorly measured branching fraction of the \lcpkpi decay, we quote the ratio
\begin{equation}
  R_{\lbdppi} \equiv \frac{\BR(\lbdppi)\times \BR(\dnkpi)}{\BR(\lblcpi)\times \BR(\lcpkpi)}\,. 
\end{equation}
The \Dz meson is reconstructed in the 
favoured final state $\Km\pip$ and the $\Lc$ baryon in the $\pkpi$ mode. 
In this way, the \lblcpi and \lbdppi decays
have the same final state particles, and some of the systematic uncertainties, 
in particular those related to particle identification (PID), cancel in the ratio. 
The branching fraction of the Cabibbo-suppressed \lbdpk decay mode is 
measured with respect to that of \lbdppi
\begin{equation}
  R_{\lbdpk} \equiv \frac{\BR(\lbdpk)}{\BR(\lbdppi)}\,. 
  \label{eq:ratio_lbdpk}
\end{equation}
The Cabibbo-suppressed decay \lblck is also studied. This decay has been considered in 
various analyses as a background component~\cite{LHCb-PAPER-2012-012, LHCb-CONF-2012-029}, 
but a dedicated study has not been performed so far. We measure the ratio
\begin{equation}
  R_{\lblck} \equiv \frac{\BR(\lblck)}{\BR(\lblcpi)}\,. 
\end{equation}

The heavier beauty-strange \xibn baryon can also decay into the final states \dpk and \lck
via $\bquark\to \cquark\uquarkbar\dquark$ colour-suppressed transitions. 
Previously, the \xibn baryon has only been observed in one decay mode, 
$\xibn\to \Xires_{\cquark}^+\pim$~\cite{Aaltonen:2011wd}, 
thus it is interesting to study other final states, as well as to measure its mass 
more precisely. Here we report measurements of the ratios of rates for \xibdpk, 
\begin{equation}
  R_{\xibdpk} \equiv \frac{f_{\xibn}\times \BR(\xibdpk)}{f_{\Lb}\times \BR(\lbdpk)}\,, 
\end{equation}
and \xiblck decays, 
\begin{equation}
  R_{\xiblck} \equiv \frac{\BR(\xiblck)\times \BR(\lcpkpi)}
                     {\BR(\xibdpk)\times \BR(\dnkpi)}\,, 
  \label{eq:ratio_xiblck}
\end{equation}
where $f_{\xibn}$ and $f_{\Lb}$ are the fragmentation fractions of the $\bquark$ quark to 
\xibn and \Lb baryons, respectively. 
The difference of \xibn and \Lb masses, $m_{\xibn}-m_{\Lb}$, is also measured.

\section{Detector description}
\label{sec:Detector}

The \lhcb detector~\cite{Alves:2008zz} is a single-arm forward
spectrometer covering the \mbox{pseudorapidity} range $2<\eta <5$,
designed for the study of particles containing \bquark or \cquark
quarks. The detector includes a high-precision tracking system
consisting of a silicon-strip vertex detector surrounding the $\proton\proton$
interaction region, a large-area silicon-strip detector located
upstream of a dipole magnet with a bending power of about
$4{\rm\,Tm}$, and three stations of silicon-strip detectors and straw
drift tubes placed downstream.
The combined tracking system provides a momentum measurement with
relative uncertainty that varies from 0.4\% at 5\gevc to 0.6\% at 100\gevc,
and impact parameter (IP) resolution of 20\mum for
tracks with high transverse momentum (\pt). Charged hadrons are identified
using two ring-imaging Cherenkov (RICH) detectors~\cite{LHCb-DP-2012-003}. 
Photon, electron and
hadron candidates are identified by a calorimeter system consisting of
scintillating-pad and preshower detectors, an electromagnetic
calorimeter and a hadronic calorimeter. Muons are identified by a
system composed of alternating layers of iron and multiwire
proportional chambers~\cite{LHCb-DP-2012-002}.

The trigger~\cite{LHCb-DP-2012-004} consists of a
hardware stage, based on information from the calorimeter and muon
systems, followed by a software stage, which applies a full event
reconstruction.
Events used in this analysis are required to satisfy at least one
hardware trigger requirement: a final state 
particle has to deposit energy in the calorimeter system above a certain 
threshold, or the event has to be triggered by any of the requirements 
not involving the signal decay products.
The software trigger requires a two-, three-, or four-track
secondary vertex with a high sum of \pt of
the tracks and a significant displacement from the primary $\proton\proton$
interaction vertices~(PVs). At least one track should have $\pt >
1.7\gevc$ and \chisqip with respect to any PV greater than 16, 
where \chisqip is defined as the
difference in \chisq of a given PV reconstructed with and
without the considered track. A multivariate algorithm~\cite{BBDT} is used for
the identification of secondary vertices consistent with the decay
of a \bquark hadron.

In the simulation, $\proton\proton$ collisions are generated using
\pythia~6.4~\cite{Sjostrand:2006za} with a specific \lhcb
configuration~\cite{LHCb-PROC-2010-056}.  Decays of hadronic particles
are described by \evtgen~\cite{Lange:2001uf}; the
interaction of the generated particles with the detector and its
response are implemented using the \geant
toolkit~\cite{Allison:2006ve, *Agostinelli:2002hh} as described in
Ref.~\cite{LHCb-PROC-2011-006}.

\section{Selection criteria}
\label{sec:Selections}

The analysis uses four combinations of final-state particles to form the \bquark-baryon candidates: 
\lcpi, \dppi, \lck, and \dpk. The \Dz mesons are reconstructed in the $\Km\pip$ final state, 
and \Lc baryons are reconstructed from \pkpi combinations. 
In addition, the combinations with the $\Dz$ meson of opposite flavour 
(\ie \dbppi and \dbpk with $\Dzb\to\Kp\pim$) are selected
to better constrain the shape of the combinatorial background in \dph final
states. These decay modes correspond to either 
doubly Cabibbo-suppressed decays of the $\Dz$, or to $\bquark\to\uquark$ transitions 
in the \Lb and \xibn decays, 
and are expected to contribute a negligible amount of signal in the current data sample. 

The selection of \bquark-baryon candidates is performed in two stages: the preselection and the final selection. 
The preselection is performed to select events containing a beauty hadron candidate 
with an intermediate charm state.
It requires that the tracks forming the candidate, as well as the 
beauty and charm vertices, have good quality and are well separated from any 
PV, and the invariant masses of the beauty and charm hadrons 
are in the region of the known values of the masses of the corresponding particles. 
The preselection has an efficiency 95--99\% for the signal depending on the decay mode. 

Two different sets of requirements are used for the final selection. 
The ratio $R_{\lbdppi}$ is measured by fitting 
the invariant mass distribution for candidates obtained with a {\it loose} selection 
to minimise the systematic uncertainty. The signal yields of these decays are 
large and the uncertainty in the ratio is dominated by systematic effects. 
The ratios $R_{\lbdpk}$ and $R_{\lblck}$ are less affected by systematic uncertainties 
since the topologies of the decays are the same. 
A {\it tight} multivariate selection is used in addition to the 
loose selection requirements when measuring these ratios, as well as the ratios of 
the \xibn decay rates.

The loose selection requires 
that the invariant masses of the intermediate \Lc and \Dz candidates are 
within 25\mevcc of their known masses~\cite{PDG2012}, and the decay time significance 
of the \Dz meson from the \lbdppi decay is greater than one standard deviation. 
The decay time significance is defined as the measured decay time divided by 
its uncertainty for a given candidate. 
The final-state particles are required to satisfy PID criteria based on 
information from the RICH detectors~\cite{LHCb-DP-2012-003}. Pion candidates are 
required to have a value $\dllkpi<5$ for the difference of logarithms of likelihoods between the 
kaon and pion hypotheses; the efficiency of this requirement is about 95\%.  
The requirement for kaon candidates of $\dllkpi>0$ is about 97\% efficient. 
The protons are required to satisfy $\dllppi>5$ and $\dllpk>0$. 
The corresponding efficiency is approximately 88\%. The momentum of each final-state track 
is required to be less than 100\gevc, corresponding to the range of good separation 
between particle types.

For candidates passing the above selections, a kinematic fit is performed~\cite{Hulsbergen:2005pu}. 
The fit employs constraints on the decay products of the $\Lb$, $\Lc$, and $\Dz$ particles 
to originate from their respective vertices, the $\Lb$ candidate to originate from the PV, and the 
\Lc and \Dz invariant masses to be equal to their known values~\cite{PDG2012}. 
A momentum scale correction is applied in the kinematic fit to improve the mass measurement
as described in Ref.~\cite{LHCb-PAPER-2012-048}. 
The momentum scale of the detector has been calibrated using inclusive $\jpsi\to\mumu$
decays to account for the relative momentum scale between different data taking periods, while 
the absolute calibration is performed with $\Bp\to \jpsi\Kp$ decays.

The tight selection is based on a boosted decision tree (BDT)~\cite{Breiman} 
trained with the gradient boost algorithm. 
The \dph selection is optimised using simulated \dpk signal events, 
and combinations with opposite-flavour \Dz candidates (\dbpk) in data as a background estimate.
The optimisation of the \lch selection is performed with a 
similar approach, with the $\Lc\Kp$ candidates as the background training sample. 
The optimisation criteria for the BDTs are the maximum expected statistical significances 
of the \lbdpk and \lblck signals, $S_{\rm stat}=N_{\rm sig}/\sqrt{N_{\rm sig} + N_{\rm bck}}$, 
where $N_{\rm sig}$ and $N_{\rm bck}$ are the expected numbers of signal and background events. 
The expected number of events for the optimisation is 
taken from the observed yields in the \lblcpi and \lbdppi modes scaled by the Cabibbo suppression 
factor. 
The variables that enter the BDT selection are the following: the quality of the kinematic fit 
($\chi^2_{\rm fit}/{\rm ndf}$, 
where ${\rm ndf}$ is the number of degrees of freedom in the fit); the minimum IP 
significance \chisqip of the final-state and intermediate charm particles
with respect to any PV; the lifetime significances of the \Lb and intermediate charm particles; 
and the PID variables (\dllppi and \dllpk) for the proton candidate. 
The \dph selection has a signal efficiency of 72\% on candidates passing the loose 
selection while retaining 11\% of the combinatorial 
background. The \lch selection is 99.5\% efficient and retains 65\% of the combinatorial 
background.

In approximately 2\% of events more than one candidate passes the 
selection. In these cases, only the candidate with the minimum $\chi_{\rm fit}^2/{\rm ndf}$
is retained for further analysis. 

Several vetoes are applied for both the loose and tight selections to reduce backgrounds. 
To veto candidates formed from $J/\psi\to\mumu$ combined with two tracks, at least one of the 
pion candidates in \lcpi and \dppi combinations is required not to have hits in the muon chambers. 
For \dph combinations, a $\Lc\to \proton\pip h^-$ veto is applied: the invariant 
mass of the $\proton\pip h^-$ combination is required to 
differ from the nominal \Lc mass by more than 20\mevcc. 
This requirement rejects the background from \lblck decays. Cross-feed between 
\lbdph and \lblcpi decays does not occur since the invariant mass of the $\Dz\proton$ combination 
in \lbdph decays is greater than the \Lc invariant mass.

\section{Determination of signal yields}
\label{sec:yields}

The signal yields are obtained from extended maximum likelihood fits to the unbinned 
invariant mass distributions. The fit model 
includes signal components (\Lb only for \lcpi and \dppi final states, and both \Lb and \xibn for 
\dpk and \lck final states), as well as various background contributions. The ratio $R_{\lbdppi}$
is obtained from the combined fit of the \lcpi and \dppi invariant mass distributions of 
candidates that pass the loose selection, while the 
other quantities are determined from the simultaneous fit of the 
\lch, \dph, and \dbph ($h=\pion$ or $\kaon$) invariant mass distributions 
passing the tight BDT-based selection requirements. 

The shape of each signal contribution is taken from simulation and is parametrised using 
the sum of two Crystal Ball (CB) functions~\cite{Skwarnicki:1986xj}. In the fit to data, 
the widths of each signal component are multiplied by a common scaling factor that is left free. 
This accounts for the difference between the invariant mass resolution observed 
in data and simulation. The masses of the \Lb and \xibn states are also free parameters. 
Their mean values as reconstructed in the \dph and \lch spectra are allowed to differ by an amount $\Delta M$
(which is the same for \Lb and \xibn masses) to account for possible imperfect calibration 
of the momentum scale in the detector. 
The mass difference $\Delta M$ obtained from the fit is consistent with zero.  

The background components considered in the analysis are subdivided into three classes: 
random combinations of tracks, or genuine \Dz or \Lc decays combined with random tracks 
(combinatorial background); decays where one or more particles are incorrectly identified
(misidentification background); 
and decays where one or more particles are not reconstructed (partially reconstructed 
background). 

The combinatorial background is parametrised with a quadratic function. 
The shapes are constrained to be the same for the \dph signal 
and \dbph background combinations.
The \dbppi fit model includes only the combinatorial background component, while 
in the \dbpk model, the \lbdbpk signal and partially reconstructed background
are included with varying yields to avoid biasing the combinatorial background shape. 
The two contributions are found to be consistent with zero, as expected. 

Contributions of charmed \B decays with misidentified particles are studied using 
simulated samples. The $\Bsb\to\Ds h^-$ and $\Bzb\to \Dp h^-$ decay modes 
are considered as \lch backgrounds, while $\Bzb\to \Dz\pip\pim$, 
$\Bzb\to \Dz\Kp\Km$~\cite{LHCb-PAPER-2012-018}, 
and $\Bsb\to \Dz\Kp\pim$~\cite{LHCb-PAPER-2013-022} are possible backgrounds in the \dph spectra. 
These contributions to \dph modes are found to be negligible and thus are not included
in the fit model, while the $\Bzb_{(s)}\to \Dp_{(s)}\pim$ component is significant 
and is included in the fit. The ratio between $\Bsb\to\Ds \pim$ and $\Bzb\to \Dp \pim$
contributions is fixed from the measured ratio of their event yields~\cite{LHCb-PAPER-2012-037}.

Contributions to \dpk and \lck spectra from the \lbdppi and \lblcpi modes, 
respectively, with the pion misidentified as a kaon ($\kaon/\pion$ misidentification
backgrounds) are obtained by parametrising the simulated samples with a CB function. 
In the case of the \lbdppi background, the 
squared invariant mass of the $\Dz\proton$ combination, $M^2(\Dz\proton)$,
is required to be smaller than $10\gevgevcccc$.
This accounts for  the dominance of events with low $\Dz\proton$ invariant 
masses observed in data. In the case of the \lcpi spectrum, the \lblck contribution 
with the kaon misidentified as a pion is also included. In all cases, the nominal 
selection requirements, including those for PID, are applied to the simulated samples. 

Partially reconstructed backgrounds, such as $\Lb\to \Dstarz\proton\pim$, $\Dstarz\to \Dz\,\piz/\g$
decays, or $\Lb\to\Sigmares_{\cquark}^+\pim$, $\Sigmares_{\cquark}^+\to \Lc\piz$ decays, 
contribute at low invariant mass. Simulation is used to check that these backgrounds are 
well separated from the signal region. However, their mass distribution 
is expected to depend strongly on the unknown helicity structure of these 
decays. Therefore, an empirical probability density function (PDF), a bifurcated Gaussian 
distribution with free parameters, is used to parametrise them. The shapes of 
the backgrounds are constrained to be the same for the \dpk and \dppi decay modes, 
as well as for the \lck and \lcpi decay modes.  

Backgrounds from partially reconstructed $\Lb\to \Dstarz\proton\pim$
and $\Lb\to\Sigmares_{\cquark}^+\pim$ decays with the pion misidentified as a kaon contribute to 
the \dpk and \lck mass spectra, respectively. These backgrounds are parametrised with 
CB functions fitted to samples simulated assuming that the amplitude is 
constant across the phase space. Their yields are constrained from the 
yields of partially reconstructed components in the \dppi and \lcpi spectra
taking into account the $\kaon/\pion$ misidentification probability.

Charmless $\Lb\to\proton\Km\pip h^-$ backgrounds, which have the same final state as the 
signal modes but no intermediate charm vertex, are studied with the \Lb invariant mass fit to data from 
the sidebands of the \dnkpi invariant mass distribution: $50<|M(\Km\pip)-m_{\Dz}|<100$\mevcc. 
Similar sidebands are used in the \lcpkpi invariant mass. 
A significant contribution is observed in the \dppi mode. Hence, for 
the \dph combinations, the \Dz vertex is required to be 
downstream of \Lb vertex and the \Dz decay time must 
differ from zero by more than one standard deviation. 
The remaining contribution is estimated from the \Lb invariant mass fit in 
the sidebands. The \lbdppi yield 
obtained from the fit is corrected for a small residual charmless contribution, 
while in other modes the contribution of this background is consistent with zero. 

\begin{figure}[t]
  \includegraphics[width=0.48\textwidth]{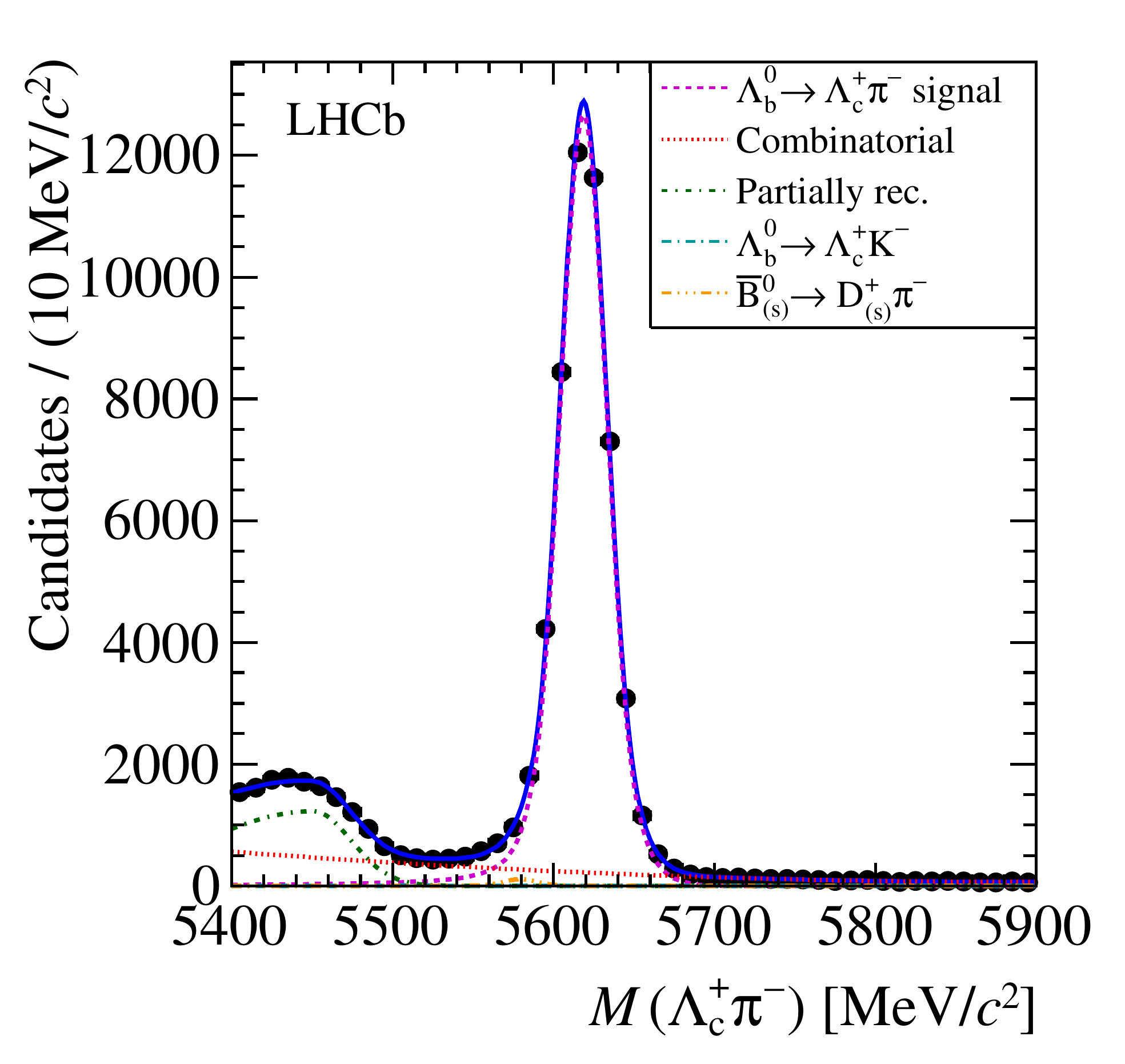}
  \put(-165,165){(a)}
  \hfill
  \includegraphics[width=0.48\textwidth]{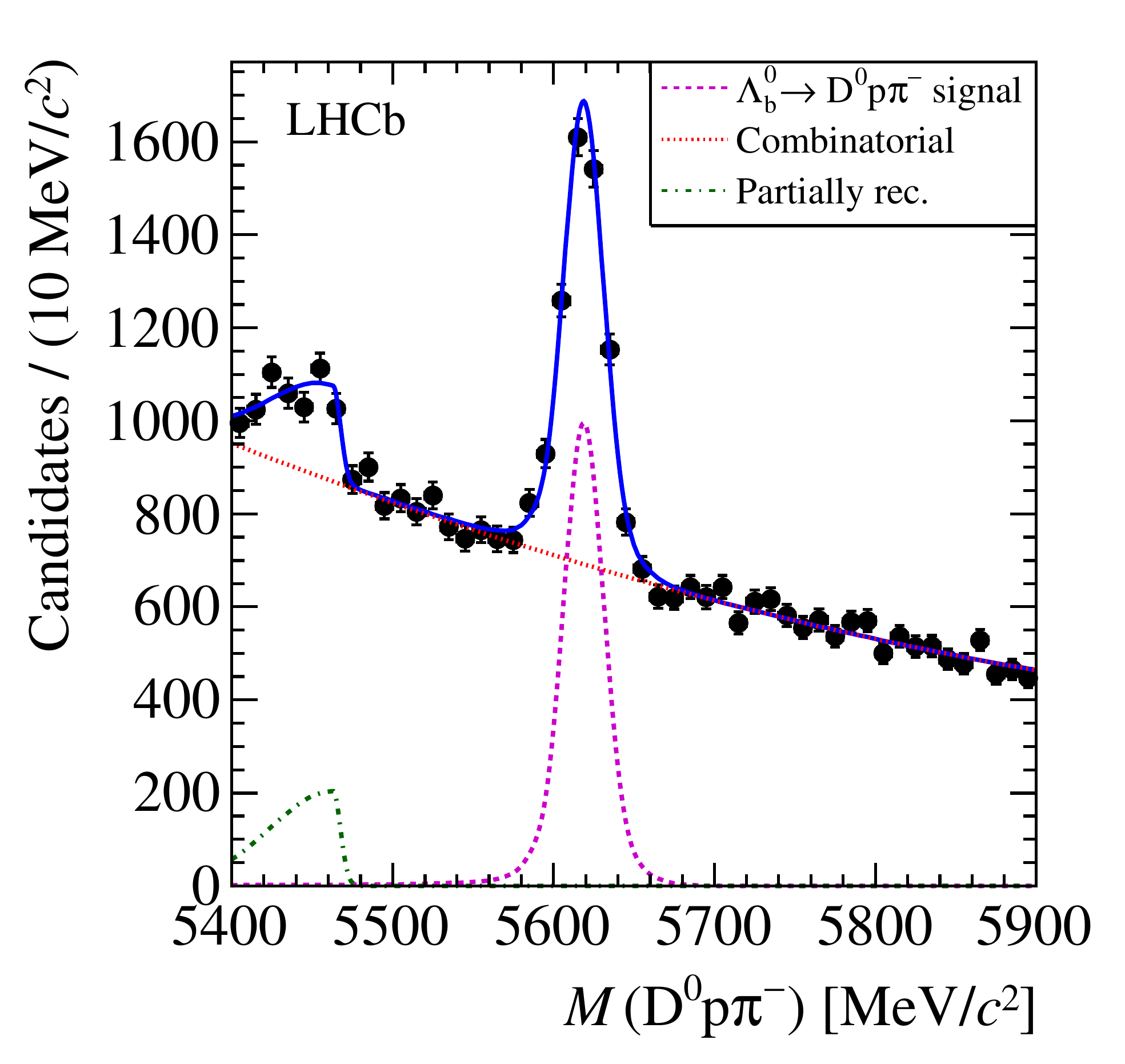}
  \put(-165,165){(b)}
  \caption{\small Distributions of invariant mass for (a) \lcpi and (b) \dppi 
           candidates passing the loose selection (points with error bars)
           and results of the fit (solid line). 
           The signal and background contributions are shown. }
  \label{fig:loose}
\end{figure}

\begin{figure}[ht!]
  \includegraphics[width=0.48\textwidth]{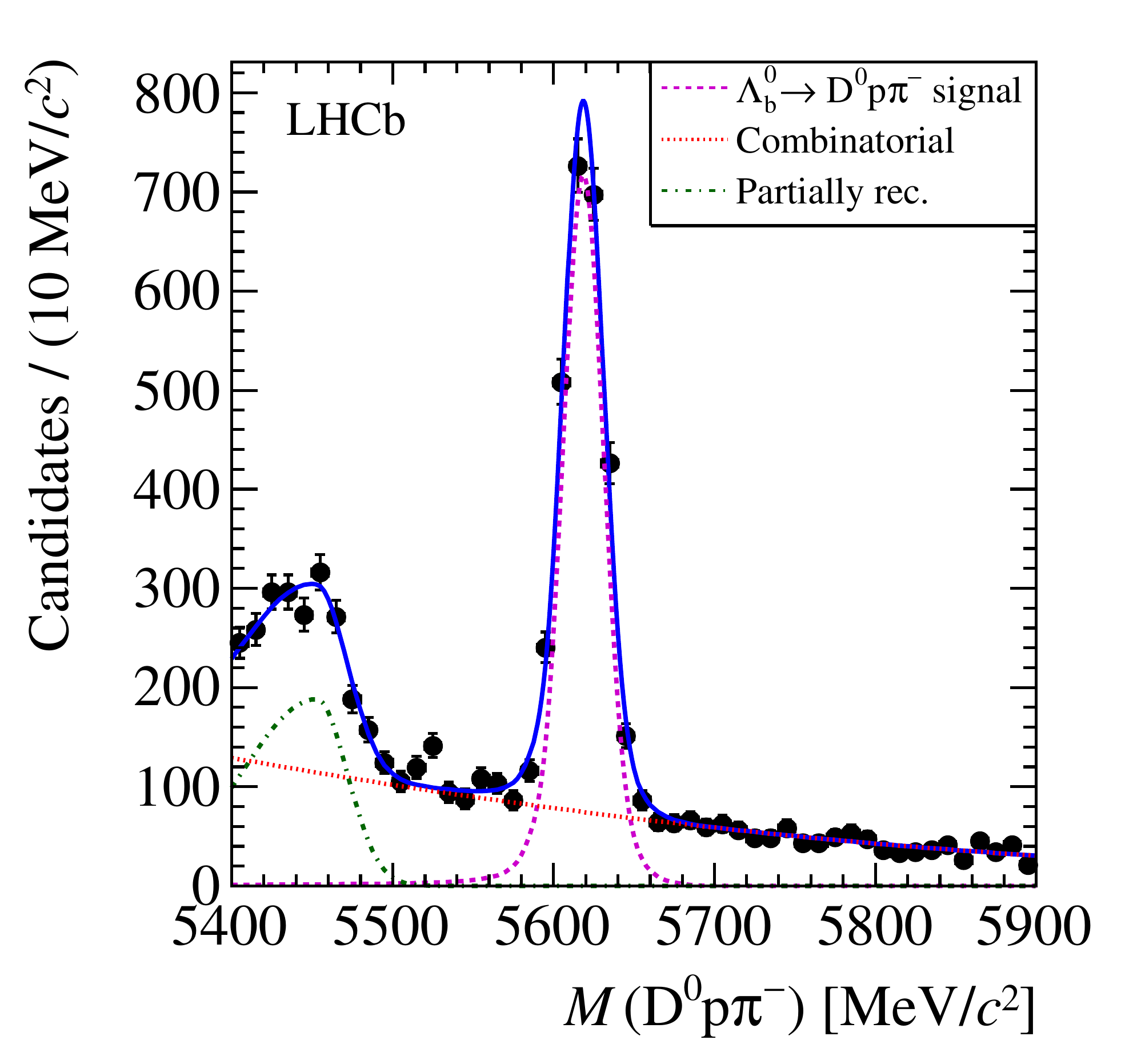}
  \put(-165,165){(a)}
  \hfill
  \includegraphics[width=0.48\textwidth]{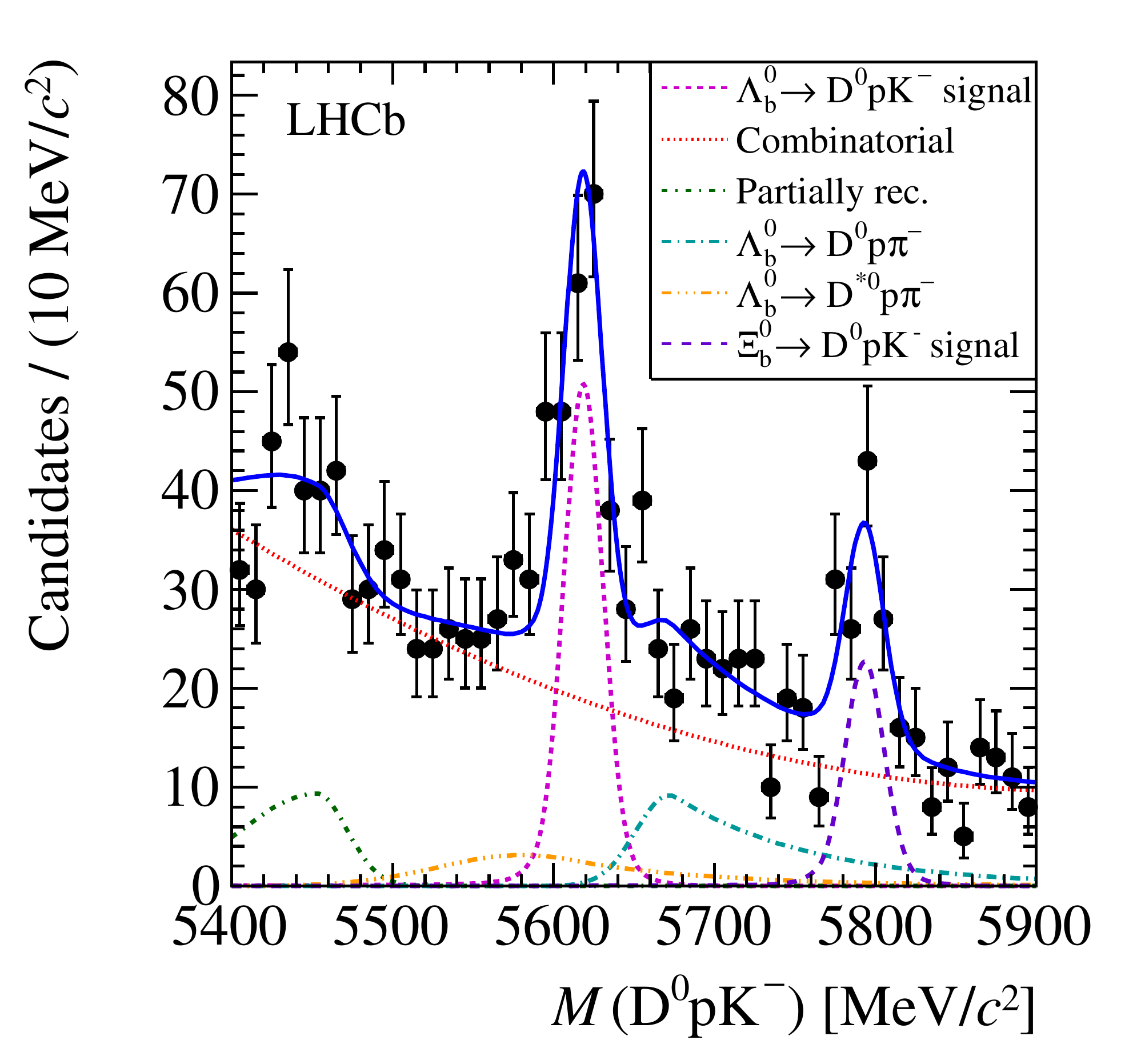}
  \put(-165,165){(b)}
  \caption{\small Distributions of invariant mass for (a) \dppi and (b) \dpk candidates passing
           the tight selection (points with error bars)
           and results of the fit (solid line). The signal and background contributions 
           are shown. }
  \label{fig:dph}
\end{figure}

\begin{figure}[ht!]
  \includegraphics[width=0.48\textwidth]{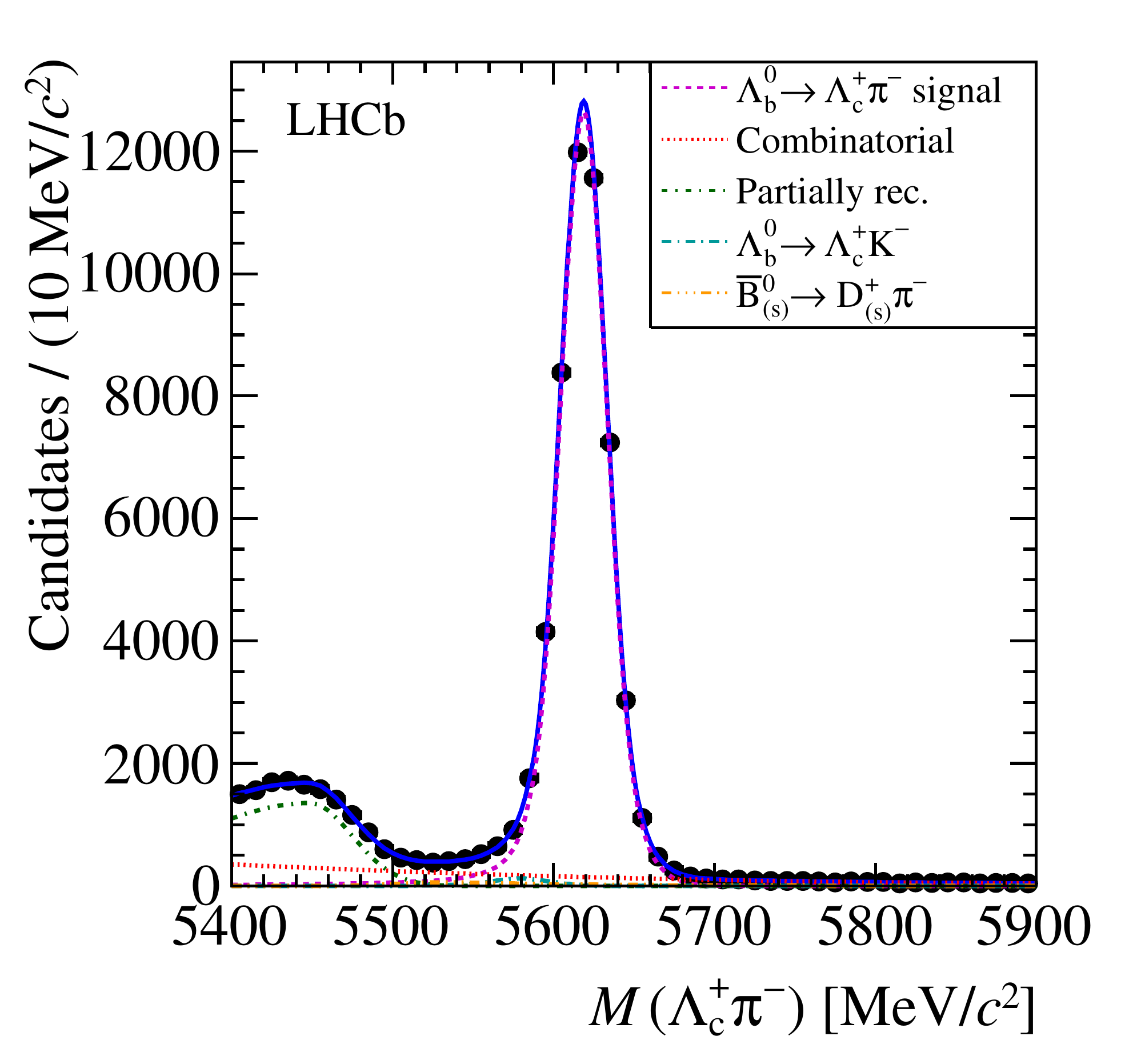}
  \put(-165,165){(a)}
  \hfill
  \includegraphics[width=0.48\textwidth]{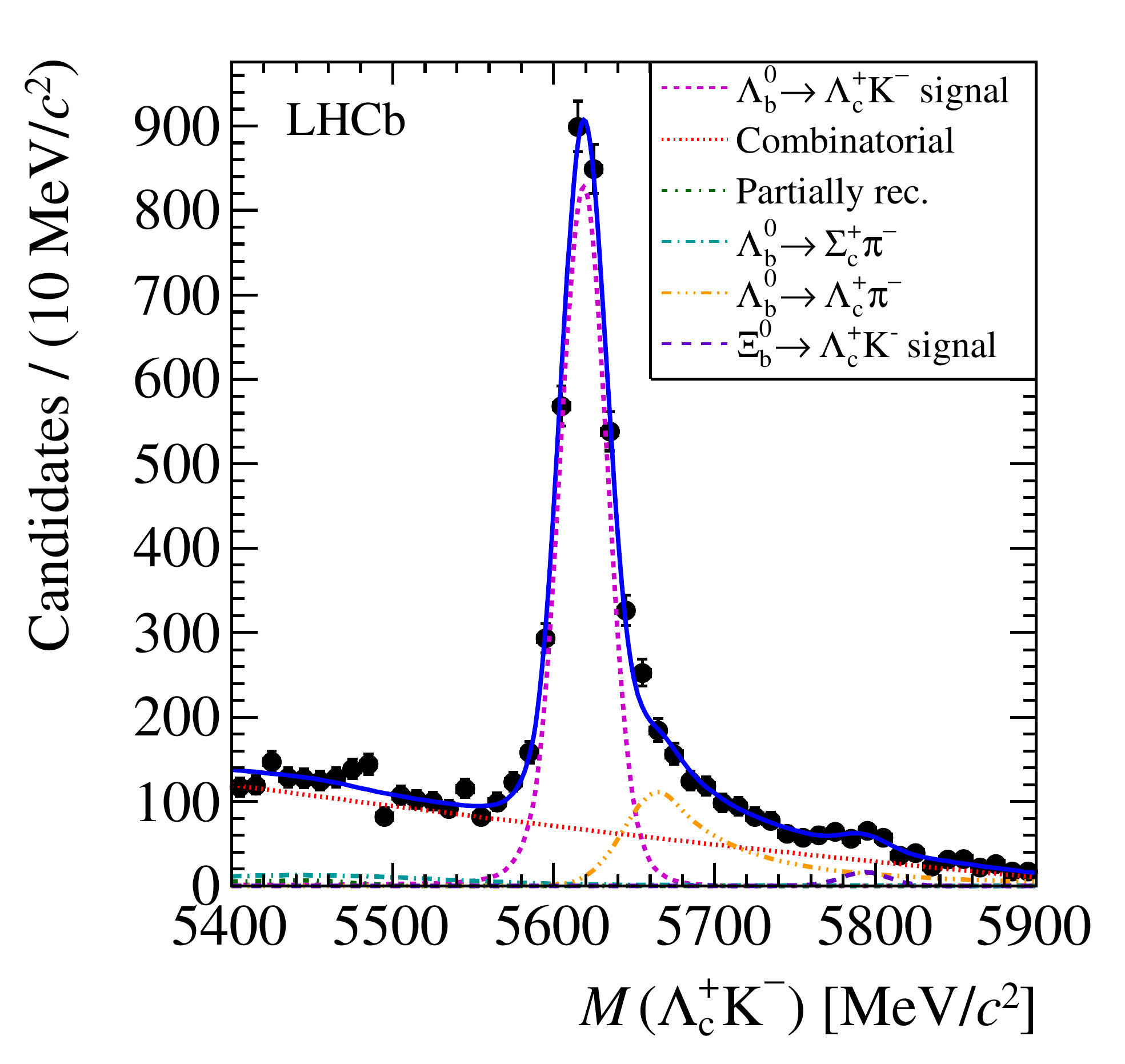}
  \put(-165,165){(b)}\\
  \includegraphics[width=0.48\textwidth]{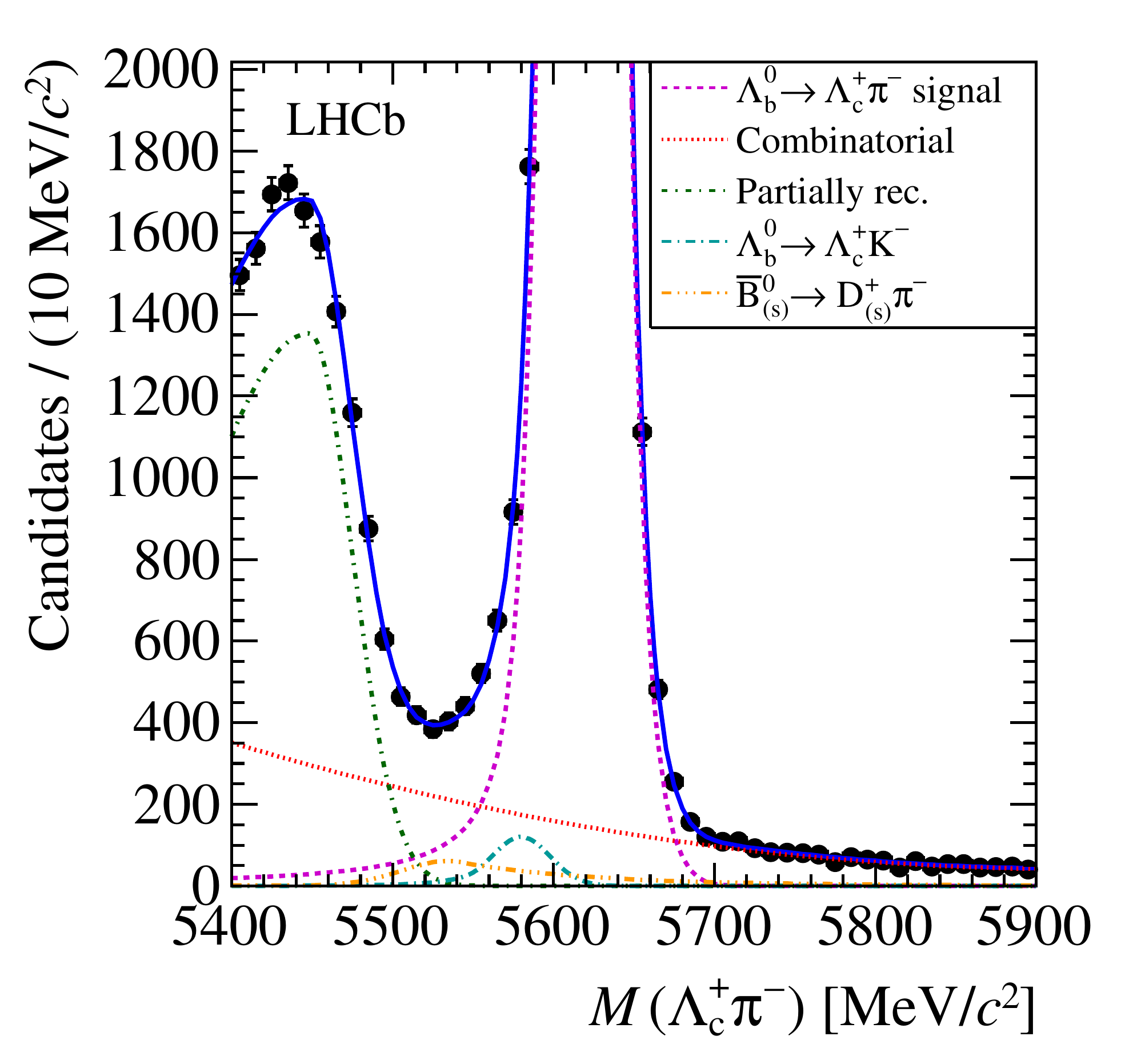}
  \put(-165,50){(c)}
  \hfill
  \includegraphics[width=0.48\textwidth]{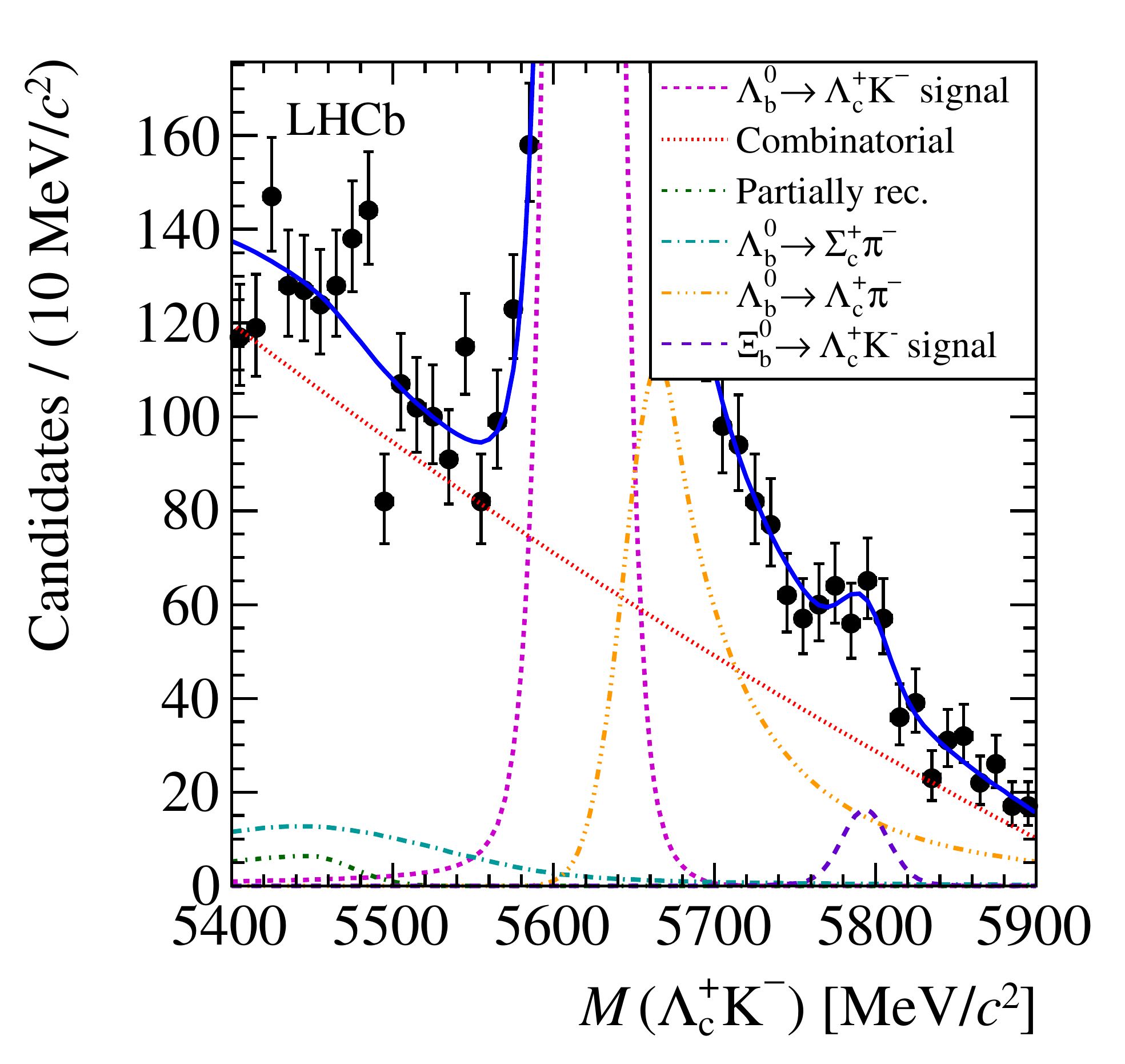}
  \put(-165,50){(d)}
  \caption{\small Distributions of invariant mass for (a) \lcpi and (b) \lck candidates 
           passing the tight selection (points with error bars)
           and results of the fit (solid line). The signal and background contributions 
           are shown. The same distributions 
           are magnified in (c) and (d) to better distinguish background components 
           and \xiblck signal. }
  \label{fig:lch}
\end{figure}

The \lcpi and \dppi invariant mass distributions obtained with the loose selection are shown in Fig.~\ref{fig:loose}
with the fit result overlaid. 
The \Lb yields obtained from the fit to these spectra are presented in Table~\ref{tab:loose_fit}. 
Figures~\ref{fig:dph} and \ref{fig:lch} show the invariant mass distributions 
for the \dph and \lch modes after the tight BDT-based selection. The \Lb and \xibn yields, 
as well as their masses, obtained from the fit are given in Table~\ref{tab:tight_fit}. 
The raw masses obtained in the fit are used to calculate the difference of \xibn and \Lb masses, 
$m_{\xibn}-m_{\Lb}=174.8\pm 2.3\mevcc$, which is less affected by the systematic uncertainty due to knowledge 
of the absolute mass scale. 

\begin{table}[t]
  \caption{\small 
           Results of the fit to the invariant mass distribution of \lblcpi and \lbdppi 
           candidates passing the loose selection. The uncertainties are statistical only. }
  \label{tab:loose_fit}
  \begin{center}
  \begin{tabular}{lc}
    \hline
    \hline
    Decay mode & Yield \\
    \hline
     \lbdppi$\vphantom{D^{0^0}}$ & \phantom{$\,$}$3383\pm 94$ \\
     \lblcpi                     &            $50\,301\pm 253$ \\
    \hline
    \hline
  \end{tabular}
  \end{center}
\end{table}

\begin{table}[t]
  \caption{\small 
           Results of the fit to the invariant mass distributions of \lblch and \lbdph candidates
           passing the tight selection. The uncertainties are statistical only. }
  \label{tab:tight_fit}
  \begin{center}
  \begin{tabular}{lc}
    \hline
    \hline
     Decay mode & Yield \\
    \hline
     \lbdppi$\vphantom{D^{0^0}}$  & \phantom{$\,00$}$2452\pm 58$\phantom{$0$} \\
     \lblcpi                      & \phantom{$0$}$50\,072\pm 253$ \\
     \lbdpk                       & \phantom{$\,000$}$163\pm 18$\phantom{$0$} \\
     \lblck                       & \phantom{$\,00$}$3182\pm 66$\phantom{$0$} \\
     \xibdpk                      & \phantom{$\,0000$}$74\pm 13$\phantom{$0$} \\
     \xiblck                      & \phantom{$\,0000$}$62\pm 20$\phantom{$0$} \\
    \hline
     Particle  & Mass $[\mevcc]\vphantom{D^{0^0}}$  \\
    \hline
     \Lb$\vphantom{D^{0^0}}$       & $5618.7\pm 0.1$ \\
     \xibn                         & $5793.5\pm 2.3$ \\
    \hline
    \hline
  \end{tabular}
  \end{center}
\end{table}

\begin{figure}[ht!]
  \includegraphics[width=\textwidth]{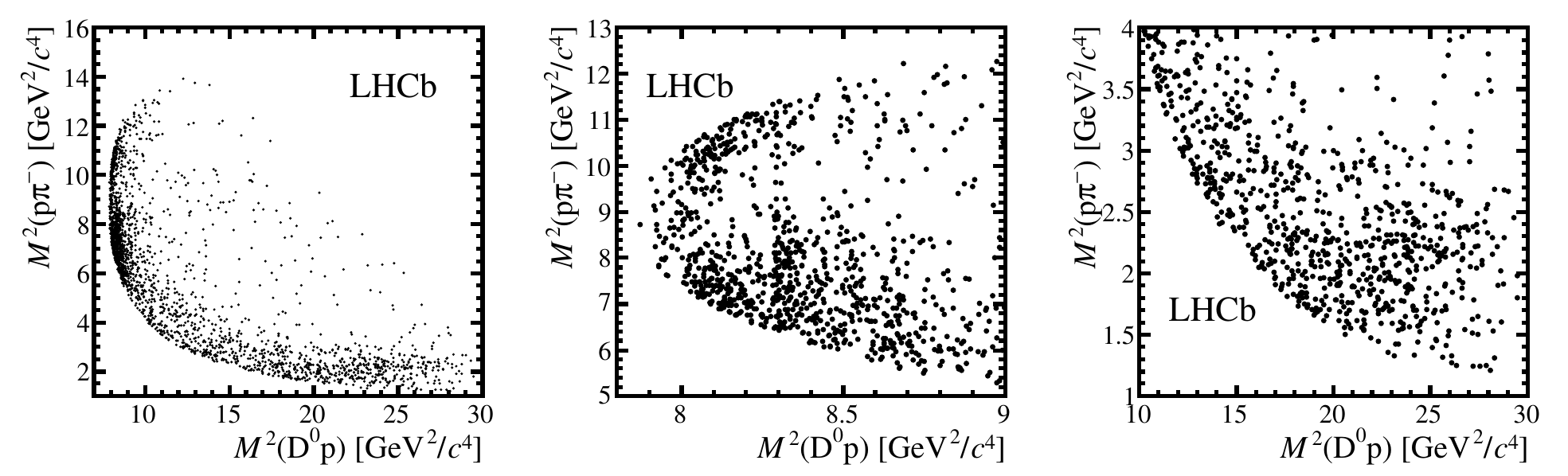}
  \put(-420,30){(a)}
  \put(-268,30){(b)}
  \put(-116,30){(c)}
  \caption{\small Dalitz plot of \lbdppi candidates in (a) the full phase space region, and magnified 
           regions of (b) low $M^2(\Dz\proton)$ and (c) low $M^2(\proton\pim)$. }
  \label{fig:dppi_dalitz}
\end{figure}

\begin{figure}[ht!]
  \begin{center}
  \includegraphics[width=0.8\textwidth]{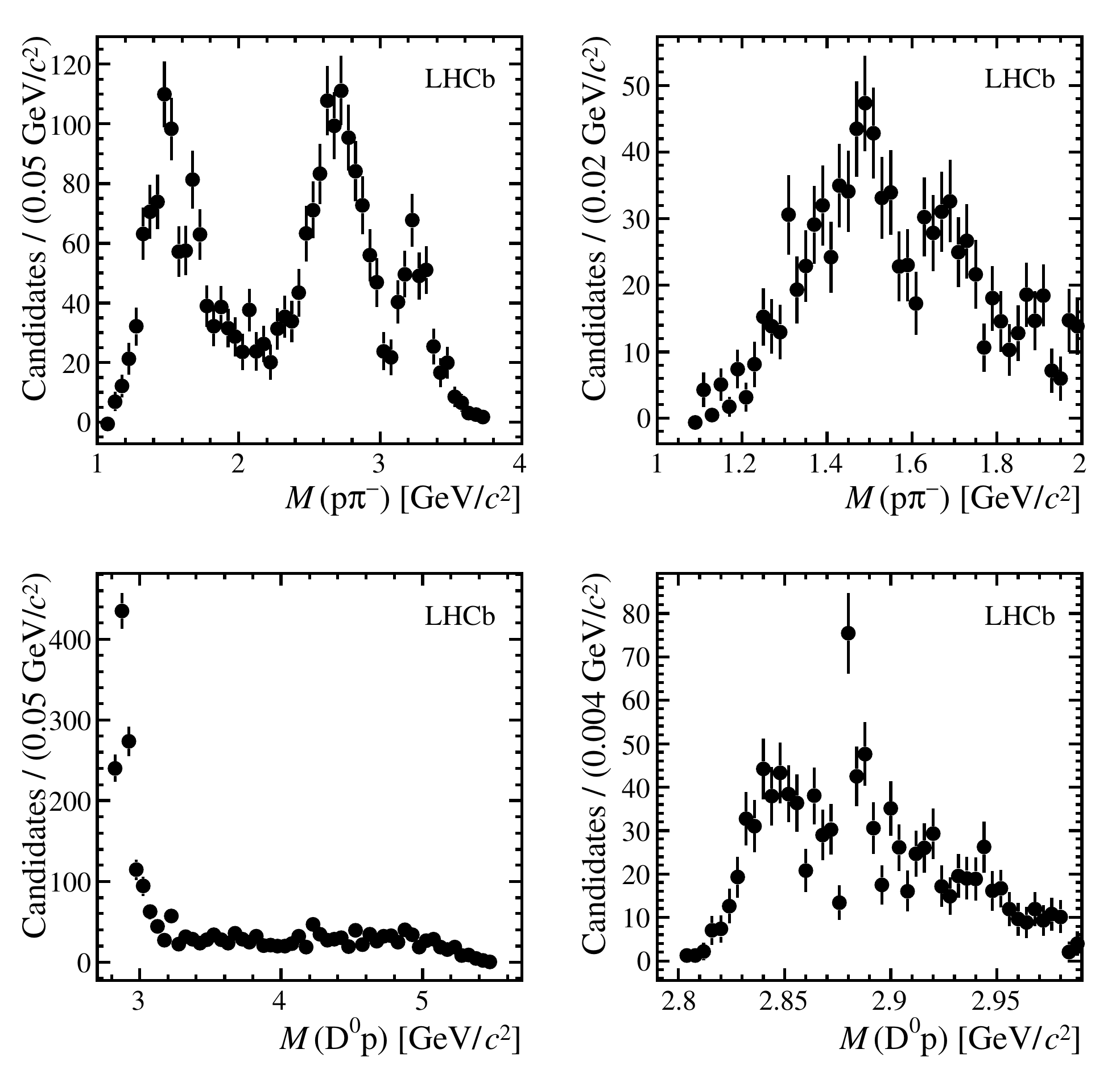}
  \put(-220,305){(a)}
  \put(-38,305){(b)}
  \put(-220,130){(c)}
  \put(-38,130){(d)}
  \end{center}
  \caption{\small Background-subtracted distributions of (a,b) $M(\proton\pim)$ 
           and (c,d) $M(\Dz\proton)$ invariant masses in \lbdppi decays, where (b) and (d) are versions of 
           (a) and (c), respectively, showing the lower invariant mass parts of the 
           distributions. The distributions are not corrected for efficiency. }
  \label{fig:dppi_dalitz_proj}
\end{figure}

\begin{figure}[ht!]
  \includegraphics[width=\textwidth]{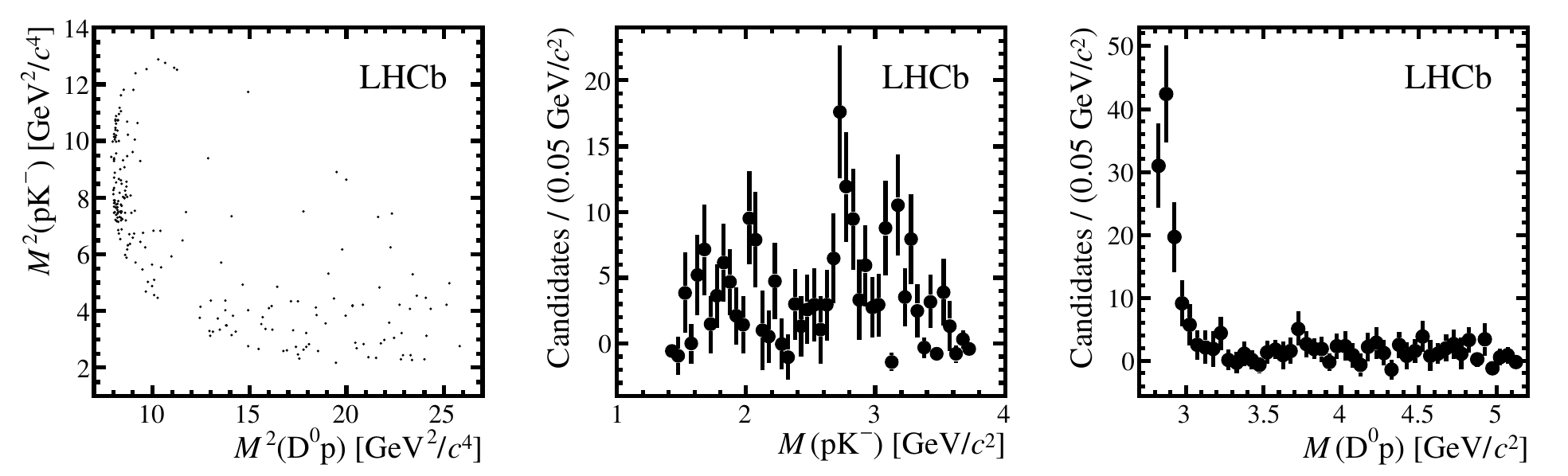}
  \put(-339,95){(a)}
  \put(-187,95){(b)}
  \put(-35 ,95){(c)}
  \caption{\small (a) \lbdpk Dalitz plot and background-subtracted distributions of 
           (b) $M(\proton\Km)$ and (c) $M(\Dz\proton)$ invariant masses. 
           The distributions are not corrected for efficiency. }
  \label{fig:dpk_dalitz}
\end{figure}

Figures~\ref{fig:dppi_dalitz} and \ref{fig:dppi_dalitz_proj} show the Dalitz plot of the 
three-body decay \lbdppi, and the projections of the two invariant masses, where resonant 
contributions are expected. 
In the projections, 
the background is subtracted using the \sPlot\ technique~\cite{Pivk:2004ty}. 
The distributions show an increased density of events in the low-$M(\Dz\proton)$ region
where a contribution from excited \Lc states is expected. 
The $\Lz_{\cquark}(2880)^+$ state is apparent in this projection. 
Structures in the $\proton\pim$ combinations are also visible. 
The Dalitz plot and projections of $\Dz\proton$ and $\proton\Km$ invariant masses 
for the \lbdpk mode are shown in Fig.~\ref{fig:dpk_dalitz}. The distributions for 
the \lbdpk mode exhibit similar behaviour with the dominance of a low-$M(\Dz\proton)$
contribution and an enhancement in the low-$M(\proton\Km)$ region. 

\section{Calculation of branching fractions}
\label{sec:branchings}

The ratios of branching fractions are calculated from the ratios of yields 
of the corresponding decays after applying several correction factors
\begin{equation}
  R = \frac{N^i}{N^j}      \frac{\varepsilon^{j}_{\rm sel}}{\varepsilon^{i}_{\rm sel}}
                           \frac{\varepsilon^{j}_{\rm PID}}{\varepsilon^{i}_{\rm PID}}
                           \frac{\varepsilon^{j}_{\rm PS}}{\varepsilon^{i}_{\rm PS}}, 
\end{equation}
where $N^{i}$ is the yield for the $i^{\mathrm{th}}$ decay mode, $\varepsilon^{i}_{\rm sel}$
is its selection efficiency excluding the PID efficiency, 
$\varepsilon^{i}_{\rm PID}$ is the efficiency of the PID requirements, and 
$\varepsilon^{i}_{\rm PS}$ is the phase-space acceptance correction defined below. 

The trigger, preselection and final selection efficiencies that enter $\varepsilon_{\rm sel}$
are obtained using simulated signal samples. The selection efficiency is calculated without the PID 
requirements applied, except for the proton PID in the tight selection, which enters the
multivariate discriminant. 
Since the multiplicities of all the final states are the same,
and the kinematic distributions of the decay products are similar, the uncertainties in 
the efficiencies largely cancel in the quoted ratios of branching fractions. 

The efficiencies of PID requirements for kaons and pions are obtained with a data-driven 
procedure using a large sample of $\Dstarp\to\Dz\pip$, $\Dz\to \Km\pip$ decays. 
The calibration sample is weighted to reproduce the kinematic properties of the 
decays under study taken from simulation. 

For protons, however, the available calibration sample $\Lz\to \proton\pim$ 
does not cover the full range in momentum-pseudorapidity space that 
the protons from the signal decays populate. Thus, in the case of the calculation of 
the ratio 
of \lblcpi and \lbdppi branching fractions, the ratio of proton 
efficiencies is taken from simulation. 
For the calculation of the ratios $\BR(\lbdpk)/\BR(\lbdppi)$ and 
$\BR(\lblck)/\BR(\lblcpi)$, where the 
kinematic properties of the proton track for the decays in the numerator and 
denominator are similar, the efficiencies are taken to be equal. 

The simulated samples used to obtain the selection efficiency are generated with 
phase-space models for the three-body \lbdph and \lcpkpi decays. The three-body distributions 
in data are, however, significantly non-uniform. Therefore, the efficiency obtained from the simulation
has to be corrected for the dependence on the three-body decay kinematic properties. 
In the case of \lbdppi decays, the relative selection efficiency as a function of $\Dz\proton$
and $\proton\pim$ squared invariant masses $\varepsilon[M^2(\Dz\proton), M^2(\proton\pim)]$ 
is determined from the phase-space simulated sample and parametrised with a polynomial function 
of fourth order. The function $\varepsilon[M^2(\Dz\proton), M^2(\proton\pim)]$ is normalised 
such that its integral is unity over the kinematically allowed phase space. 
The efficiency correction factor $\varepsilon_{\rm PS}$ is calculated as
\begin{equation}
  \varepsilon_{\rm PS} = \frac{\sum_i w_i}{\sum_i w_i/\varepsilon[M_i^2(\Dz\proton), M_i^2(\proton\pim)]}, 
  \label{eq:dlz_eff}
\end{equation}
where $M^2_i(\Dz\proton)$ and $M_i^2(\proton\pim)$ are the squared invariant masses 
of the $\Dz\proton$ and $\proton\pim$ combinations for the $i^{\mathrm{th}}$ event in data, and 
$w_i$ is its signal weight obtained from the $M(\dph)$ invariant mass fit. The correction 
factor for the \lcpkpi decay is calculated similarly. 

Since the three-body decays \lcpkpi and \lbdph involve particles with non-zero spin 
in the initial and final states, the kinematic properties
of these decays are described by angular variables in addition to the two Dalitz plot variables. 
The variation of the selection efficiency with the angles can thus affect the 
measurement. We use three 
independent variables to parametrise the angular phase space, similar to those used in
Ref.~\cite{Aitala:1999uq} for the analysis of the \lcpkpi decay. 
The variables are defined in the rest frame of the decaying \Lb or \Lc baryons, 
with the $x$ axis given by their direction in the laboratory frame, the polarisation axis 
$z$ given by the cross product of the beam and $x$ axes, 
and the $y$ axis by the cross product of the $z$ and $x$ axes. The three variables are 
the cosine of the polar angle $\theta_p$ of the proton momentum in this reference frame, 
the azimuthal angle $\phi_p$ of the proton momentum in the reference frame, and 
the angle between the $\Dz h^-$-plane (for \lbdph) or $\Km\pip$-plane (for \lcpkpi) 
and the plane formed by the proton and polarisation axis. 
The angular acceptance corrections are calculated from background-subtracted angular distributions 
obtained from the data. The distributions are similar to those obtained from the simulation of 
unpolarised \Lb decays, supporting the observation of small \Lb polarisation in $\proton\proton$ 
collisions~\cite{LHCb-PAPER-2012-057}. The angular corrections are found to 
be negligible and are not used in the calculation of the ratios of branching fractions. 

\begin{table}
  \caption{\small Efficiency correction factors used to calculate the ratios of branching fractions.}
  \label{tab:effcorr}
  \begin{tabular}{lccccc}
    \hline\hline
     Correction factor & $R_{\lbdppi}$ & $R_{\lbdpk}$ & $R_{\lblck}$ & $R_{\xibdpk}$ & $R_{\xiblck}$ \\
    \hline
    $\varepsilon^i_{\rm sel}$/$\varepsilon^j_{\rm sel}$ & 
                         1.18          & 1.01         & 0.99         & 0.97          & 0.68          \\
    $\varepsilon^i_{\rm PID}$/$\varepsilon^j_{\rm PID}$ & 
                         0.98          & 1.06         & 1.17         & --            & 1.07          \\
    $\varepsilon^i_{\rm PS}$/$\varepsilon^j_{\rm PS}$ & 
                         1.03          & 1.02         & --           & --            & 0.92          \\    
    \hline\hline
  \end{tabular}
\end{table}

The values of the efficiency correction factors are given in Table~\ref{tab:effcorr}. 
The values of the branching fraction ratios defined in Eqs.~(\ref{eq:ratio_lbdpk}--\ref{eq:ratio_xiblck}) 
obtained after corrections as described above, and their statistical uncertainties, are given in 
Table~\ref{tab:syst}.

\newcommand{\pha}{\phantom{$<\;$}}
\newcommand{\phb}{\phantom{$<0$}}
\newcommand{\phc}{\phantom{$0$}}

\begin{table}
  \caption{\small Measured ratios of branching fractions, with 
           their statistical and systematic uncertainties in units of $10^{-2}$. }
  \label{tab:syst}
  \begin{tabular}{lccccc}
    \hline
    \hline
                               & $R_{\lbdppi}$\hspace{-2.1mm}
                               & $R_{\lbdpk}$\hspace{-2.1mm}
                               & $R_{\lblck}$\hspace{-2.1mm}
                               & $R_{\xibdpk}$\hspace{-2.1mm}
                               & $R_{\xiblck}$\hspace{-1.5mm} \\
    \hline
      Central value            & \pha8.06 & 7.27 & \pha7.31 & \pha44.3   & \pha57   \\
      Statistical uncertainty  & \pha0.23 & 0.82 & \pha0.16 & \phb9.2    & \pha22   \\
    \hline
      Systematic uncertainties &      &      &          &            &          \\
      ~~Signal model           & \pha0.03 & 0.03 & \pha0.05 & \phb0.2    & \phb3    \\
      ~~Background model       & \pha0.07 & ${}^{+0.34}_{-0.54}$ 
                                             & \pha0.09 & \phb5.0    & \pha20   \\
      ~~Trigger efficiency     & \pha0.01 & 0.08 & \pha0.07 & \phc$<0.1$ & \phc$<1$ \\
      ~~Reconstruction efficiency\hspace{-4.5mm} 
                               & $<0.01$  & 0.04 & \pha0.04 & \phc$<0.1$ & \phc$<1$ \\
      ~~Selection efficiency   & \pha0.12 & 0.01 & $<0.01$  & \phc$<0.1$ & \phc$<1$ \\
      ~~Simulation sample size\hspace{-4.5mm} 
                               & \pha0.06 & 0.07 & \pha0.08 & \phb0.6    & \phc$<1$ \\ 
      ~~Phase space acceptance\hspace{-4.5mm} 
                               & \pha0.07 & 0.04 & \pha--   & \phc$<0.1$ & \phc$<1$ \\
      ~~Angular acceptance     & \pha0.15 & 0.29 & \pha--   & \phb3.5    & \phb4    \\
      ~~PID efficiency         & \pha0.26 & 0.11 & \pha0.04 & \phb--     & \phb1    \\
    \hline 
      Total systematic uncertainty\hspace{-5mm}    
                               & \pha0.35 & ${}^{+0.48}_{-0.64}$\vphantom{${}^{0^{0^0}}$} 
                               & \pha0.16 & \phb6.0         & \pha21   \\
    \hline
    \hline
  \end{tabular}
\end{table}

\section{Systematic uncertainties}

The systematic uncertainties in the measurements of the ratios of 
branching fractions are listed in Table~\ref{tab:syst}. 

The uncertainties due to the description of signal and background contributions 
in the invariant mass fit model are estimated as follows: 
\begin{itemize}

\item The uncertainty due to the parametrisation of the signal distributions is obtained 
by using an alternative description based on a double-Gaussian shape, or a 
triple-Gaussian shape in the case of \lblcpi. 

\item To determine the uncertainty due to the combinatorial background parametrisation, 
an alternative model with an exponential distribution is used instead of the quadratic
polynomial function. 

\item The uncertainty in the parametrisation of the backgrounds from \B meson decays with 
misidentified particles in the final state is estimated by removing the 
$\Bzb_{(s)}\to\Dp_{(s)}\pim$ contribution. The uncertainty due to the parametrisaton of the 
$\kaon/\pion$ misidentification background is estimated by using the shapes obtained without the PID requirements
and without rejecting the events with the $\Dz\proton$ invariant mass squared greater 
than 10\gevgevcccc in the fit to the simulated sample. 

\item The uncertainty due to the partially reconstructed background is estimated 
by fitting the invariant mass distributions in the reduced range of 5500--5900\mevcc, 
and by excluding the contributions of partially reconstructed backgrounds with 
$\kaon/\pion$ misidentification from the fit for \dpk and \lck combinations. 

\item The uncertainty due to the charmless background component $\Lb\to\proton\Km\pip h^-$
is estimated from the fit of the \dph (\lch) invariant mass distributions 
in the sidebands of the \Dz (\Lc) candidate invariant mass. 
\end{itemize}

A potential source of background that is not included in the fit comes 
from \xibn baryon decays into $\Dstarz\proton\Km$ or similar final states, which 
differ from the reconstructed \dpk state by missing low-momentum particles. Such decays can 
contribute under the \lbdpk signal peak. The possible contribution of these decays is estimated 
assuming that
$\BR(\xibn\to\Dstarz\proton\Km)/\BR(\xibdpk)$ is equal to 
$\BR(\Lb\to\Dstarz\proton\Km)/\BR(\lbdpk)$
and that the selection efficiencies for \xibn and \Lb decays are the same. 
The one-sided systematic uncertainty due to this effect is added to the background 
model uncertainty for the \lbdpk decay mode.

The trigger efficiency uncertainty is dominated by the difference of the 
transverse energy threshold of the hardware-stage trigger observed between simulation and data. 
It is estimated by varying the transverse energy threshold in the simulation by 15\%. 
In the case of measuring the ratios $R_{\lbdpk}$ and $R_{\lblck}$, one also has to take into
account the difference of hadronic interaction cross section for kaons and pions 
before the calorimeter. This difference is studied using a sample of $\Bp\to \Dzb\pip$, 
$\Dzb\to \Kp\pim$ decays that pass the trigger decision independent of the final state 
particles of these decays. The difference was found to be 4.5\% for \dph and 2.5\% for \lch. 
Since only about 13\% of events are triggered exclusively by the $h^-$ particle, 
the resulting uncertainty is low.

The uncertainty due to track reconstruction efficiency cancels to a good approximation 
for the quoted ratios since the track multiplicities of the decays are the same. However, for the 
ratios $R_{\lbdpk}$ and $R_{\lblck}$, the difference in hadronic interaction rate for kaons 
and pions in the tracker can bias the measurement. A systematic uncertainty is assigned 
taking into account the rate of hadronic interactions in the simulation and the uncertainty 
on the knowledge of the amount of material in the \lhcb tracker.

The uncertainty in the selection efficiency obtained from simulation is evaluated 
by scaling the variables that enter the offline selection. The scaling factor is chosen 
from the comparison of the distributions of these variables in simulation and in a 
background-subtracted \lblcpi sample. 
In addition, the uncertainty due to the finite size of the simulation samples is assigned. 

The uncertainty of the phase-space efficiency correction includes four effects. 
The statistical uncertainty on the correction factor is determined by the 
data sample size and variations of the efficiency over the phase space. 
The uncertainty in the parametrisation of the efficiency shape is estimated by using an 
alternative parametrisation with a third-order rather than a fourth-order polynomial. The 
correlation of the efficiency shape and invariant mass of \Lb (\xibn) candidates is estimated by 
calculating the efficiency shape in three bins of \Lb (\xibn) mass separately and 
using one of the three shapes depending on the invariant mass of the candidate. 
The uncertainty due to the difference of the \Lb (\xibn) kinematic properties between 
simulation and data is estimated by using the efficiency shape obtained 
after weighting the simulated sample using the momentum 
distribution of \Lb (\xibn) from background-subtracted \lblcpi data.

Corrections due to the angular acceptance in the calculation of ratios of branching fractions 
are consistent with zero. The central values quoted do not include these corrections, 
while the systematic uncertainty is evaluated by taking the maximum of the statistical 
uncertainty for the correction, determined by the size of the data sample, and the 
deviation of its central value from unity. 

The uncertainty in the PID response is calculated differently for the ratio of \lbdppi and 
\lblcpi branching fractions using loose selection, and for the measurements using tight 
BDT-based selections. 
For the ratio of \lbdppi and \lblcpi branching fractions, $R_{\lbdppi}$, the uncertainty due 
to the pion and kaon PID requirements is estimated by scaling the PID 
variables within the limits given by the comparison of distributions from the 
reweighted calibration sample and the background-subtracted \lblcpi data. 
The dominant contribution to the PID uncertainty comes from the uncertainty in the proton PID 
efficiency ratio, which is caused by the difference in kinematic properties of the 
proton from \lbdppi and \lblcpi decays.
The proton efficiency ratio in this case is taken from simulation, 
and the systematic uncertainty is estimated by taking this ratio to be equal to one. 
In the case of measuring the ratios $R_{\lbdpk}$ and $R_{\lblck}$, the uncertainty 
due to the proton PID and the tracks coming from the \Dz or \Lc candidates is 
negligible due to similar kinematic distributions of the decays in the numerator
and denominator. The dominant contribution comes from the PID efficiency ratio 
for the kaon or pion track from the \Lb vertex; this is estimated by scaling the 
PID distribution as described above. In addition, there are contributions 
due to the finite size of the PID calibration sample, and the uncertainty due to 
assumption that the PID efficiency for the individual tracks factorises in the 
total efficiency. The latter is estimated with simulated samples. 

Since the results for the \Lb decay modes are all ratios to other \Lb decays, 
there is no systematic bias introduced by the dependence of the efficiency 
on the \Lb lifetime, and the fact that the value used in the simulation ($1.38\ps$) 
differs from the latest measurement~\cite{LHCb-PAPER-2013-032}. 
We also do not assign any systematic uncertainty due to the lack of 
knowledge of the \xibn lifetime, which is as-yet unmeasured 
(a value of $1.42\ps$ is used in the simulation).

The dominant systematic uncertainties in the measurement of the \xibn and \Lb mass difference 
(see Table~\ref{tab:syst_mass}) come from the uncertainties of the signal 
and background models, and are estimated from the 
same variations of these models as in the calculation of branching fractions. The uncertainty due to 
the momentum scale calibration partially cancels in the quoted difference of \xibn and \Lb masses; 
the residual contribution is estimated by varying the momentum scale factor 
within its uncertainty of 0.3\%~\cite{LHCb-PAPER-2012-048}. 

\begin{table}
  \caption{\small Systematic uncertainties in the measurement of the mass difference $m_{\xibn}-m_{\Lb}$.}
  \label{tab:syst_mass}
  \centering
  \begin{tabular}{lc}
    \hline\hline
    Source           & Uncertainty~(${\mathrm{Me\kern -0.1em V\!/}c^2}$) \\
    \hline
    Signal model     & 0.19 \\
    Background model & 0.50 \\
    Momentum scale calibration & 0.03 \\
    \hline
    Total            & 0.54 \\
    \hline\hline
  \end{tabular}
\end{table}

\section{Signal significance and fit validation}

The statistical significance of the \lbdpk, \xibdpk, and \xiblck signals, expressed in terms of 
equivalent number of standard deviations ($\sigma$), is evaluated from the maximum likelihood fit as 
\begin{equation}
  S_{\rm stat} = \sqrt{-2\Delta\ln\mathcal{L}}, 
  \label{eq:significance}
\end{equation}
where $\Delta\ln\mathcal{L}$ is the difference in logarithms of the likelihoods for the fits with 
and without the corresponding signal contribution. The fit yields the statistical 
significance of the \lbdpk, \xibdpk, and \xiblck signals of 
$10.8\,\sigma$, $6.7\,\sigma$, and $4.7\,\sigma$, respectively. 

The validity of this evaluation is checked with the following procedure.
To evaluate the significance of each signal, a large number of invariant mass 
distributions is generated using the result of the fit on data as input, excluding 
the signal contribution under consideration. 
Each distribution is then fitted with models 
that include background only, as well as background and signal. The significance is obtained 
as the fraction of samples where the difference $\Delta\ln\mathcal{L}$ 
for the fits with and without the signal is larger than in data. 
The significance evaluated from the likelihood fit according to 
Eq.~(\ref{eq:significance}) is consistent with, or slightly smaller than that
estimated from the simulated experiments. 
Thus, the significance calculated as in Eq.~(\ref{eq:significance}) is taken. 

The significance accounting for the systematic uncertainties is evaluated as 
\begin{equation}
  S_{\rm stat+syst} = S_{\rm stat}\left/\sqrt{1 + \sigma^2_{\rm syst}/\sigma^2_{\rm stat}}\right., 
\end{equation}
where $\sigma_{\rm stat}$ is the statistical uncertainty of the signal yield 
and $\sigma_{\rm syst}$ is the corresponding systematic uncertainty, 
which only includes the relevant uncertainties due to the signal and background models. 
As a result, the significance 
for the \lbdpk, \xibdpk, and \xiblck signals is calculated to be $9.0\,\sigma$, $5.9\,\sigma$, 
and $3.3\,\sigma$, respectively. 

The fitting procedure is tested with simulated experiments where the invariant mass distributions 
are generated from the PDFs that are a result of the data fit, 
and then fitted with the same procedure as applied to data. 
No significant biases are introduced by the fit procedure in the fitted parameters. 
However, we find that
the statistical uncertainty on the \xibn mass is underestimated by 3\% in the fit 
and the uncertainty on the \xibdpk yield is underestimated by 5\%. 
We apply the corresponding scale factors to the \xibdpk yield and \xibn mass 
uncertainties to obtain the final results. 

\section{Conclusion}
\label{sec:Conclusion}

We report studies of beauty baryon decays to the \dph and \lch final states, 
using a data sample corresponding to an integrated luminosity of 1.0\invfb collected with the LHCb detector.
First observations of the \lbdpk and \xibdpk decays are reported, with 
significances of 9.0 and 5.9 standard deviations, respectively. 
The decay \lblck is observed for the first time; the significance of this observation is 
greater than 10 standard deviations. 
The first evidence for the \xiblck decay is also obtained with a significance of 3.3 standard deviations.

The combinations of branching and fragmentation fractions for 
beauty baryons decaying into \dph and \lch final states are measured to be
\begin{equation*}
  \begin{split}
  R_{\lbdppi}\equiv\frac{\BR(\lbdppi)\times \BR(\dnkpi)}
       {\BR(\lblcpi)\times \BR(\lcpkpi)} & = 0.0806 \pm 0.0023\pm 0.0035, \\
  R_{\lbdpk}\equiv\frac{\BR(\lbdpk)}{\BR(\lbdppi)} & = 0.073 \pm 0.008\,^{+0.005}_{-0.006}, \\
  R_{\lblck}\equiv\frac{\BR(\lblck)}{\BR(\lblcpi)} & = 0.0731 \pm 0.0016\pm 0.0016, \\
  R_{\xibdpk}\equiv\frac{f_{\xibn}\times \BR(\xibdpk)}{f_{\Lb}\times \BR(\lbdpk)} & = 0.44 \pm 0.09\pm 0.06, \\
  R_{\xiblck}\equiv\frac{\BR(\xiblck)\times \BR(\lcpkpi)}
       {\BR(\xibdpk)\times \BR(\dnkpi)} & = 0.57 \pm 0.22\pm 0.21, \\
  \end{split}
\end{equation*}
where the first uncertainty is statistical and the second systematic. 
The ratios of the Cabibbo-suppressed to Cabibbo-favoured branching fractions for both the \dph and 
the \lch modes are consistent with the those observed for the $\B\to\D h$ modes~\cite{PDG2012}. 
In addition, the difference of \xibn and \Lb baryon masses is measured to be 
\[
  m_{\xibn}-m_{\Lb}=174.8\pm 2.4\pm 0.5\mevcc. 
\]
Using the latest LHCb measurement of the \Lb mass 
$m_{\Lb}=5619.53\pm 0.13\pm 0.45\mevcc$~\cite{LHCb-PAPER-2012-048}, 
the \xibn mass is determined to be $m_{\xibn}=5794.3\pm 2.4\pm 0.7$\mevcc,
in agreement with the measurement performed by CDF~\cite{Aaltonen:2011wd} and twice as precise.

\section*{Acknowledgements}

\noindent We express our gratitude to our colleagues in the CERN
accelerator departments for the excellent performance of the LHC. We
thank the technical and administrative staff at the LHCb
institutes. We acknowledge support from CERN and from the national
agencies: CAPES, CNPq, FAPERJ and FINEP (Brazil); NSFC (China);
CNRS/IN2P3 and Region Auvergne (France); BMBF, DFG, HGF and MPG
(Germany); SFI (Ireland); INFN (Italy); FOM and NWO (The Netherlands);
SCSR (Poland); MEN/IFA (Romania); MinES, Rosatom, RFBR and NRC
``Kurchatov Institute'' (Russia); MinECo, XuntaGal and GENCAT (Spain);
SNSF and SER (Switzerland); NAS Ukraine (Ukraine); STFC (United
Kingdom); NSF (USA). We also acknowledge the support received from the
ERC under FP7. The Tier1 computing centres are supported by IN2P3
(France), KIT and BMBF (Germany), INFN (Italy), NWO and SURF (The
Netherlands), PIC (Spain), GridPP (United Kingdom). We are thankful
for the computing resources put at our disposal by Yandex LLC
(Russia), as well as to the communities behind the multiple open
source software packages that we depend on.

\addcontentsline{toc}{section}{References}
\setboolean{inbibliography}{true}
\bibliographystyle{LHCb}
%\bibliography{main,LHCb-PAPER,LHCb-CONF,LHCb-DP}
\bibliography{main}

\ifx\mcitethebibliography\mciteundefinedmacro
\PackageError{LHCb.bst}{mciteplus.sty has not been loaded}
{This bibstyle requires the use of the mciteplus package.}\fi
\providecommand{\href}[2]{#2}
\begin{mcitethebibliography}{10}
\mciteSetBstSublistMode{n}
\mciteSetBstMaxWidthForm{subitem}{\alph{mcitesubitemcount})}
\mciteSetBstSublistLabelBeginEnd{\mcitemaxwidthsubitemform\space}
{\relax}{\relax}

\bibitem{PDG2012}
Particle Data Group, J.~Beringer {\em et~al.},
  \ifthenelse{\boolean{articletitles}}{{\it {\href{http://pdg.lbl.gov/}{Review
  of particle physics}}},
  }{}\href{http://dx.doi.org/10.1103/PhysRevD.86.010001}{Phys.\ Rev.\  {\bf
  D86} (2012) 010001}, and 2013 partial update for the 2014 edition.\relax
\mciteBstWouldAddEndPunctfalse
\mciteSetBstMidEndSepPunct{\mcitedefaultmidpunct}
{}{\mcitedefaultseppunct}\relax
\EndOfBibitem
\bibitem{:2007rw}
CDF collaboration, T.~Aaltonen {\em et~al.},
  \ifthenelse{\boolean{articletitles}}{{\it {Observation of the heavy baryons
  $\Sigmares_b$ and $\Sigmares_b^*$}},
  }{}\href{http://dx.doi.org/10.1103/PhysRevLett.99.202001}{Phys.\ Rev.\ Lett.\
   {\bf 99} (2007) 202001}, \href{http://arxiv.org/abs/0706.3868}{{\tt
  arXiv:0706.3868}}\relax
\mciteBstWouldAddEndPuncttrue
\mciteSetBstMidEndSepPunct{\mcitedefaultmidpunct}
{\mcitedefaultendpunct}{\mcitedefaultseppunct}\relax
\EndOfBibitem
\bibitem{Aaltonen:2011wd}
CDF collaboration, T.~Aaltonen {\em et~al.},
  \ifthenelse{\boolean{articletitles}}{{\it {Observation of the $\Xires_b^0$
  baryon}}, }{}\href{http://dx.doi.org/10.1103/PhysRevLett.107.102001}{Phys.\
  Rev.\ Lett.\  {\bf 107} (2011) 102001},
  \href{http://arxiv.org/abs/1107.4015}{{\tt arXiv:1107.4015}}\relax
\mciteBstWouldAddEndPuncttrue
\mciteSetBstMidEndSepPunct{\mcitedefaultmidpunct}
{\mcitedefaultendpunct}{\mcitedefaultseppunct}\relax
\EndOfBibitem
\bibitem{Dunietz:1992ti}
I.~Dunietz, \ifthenelse{\boolean{articletitles}}{{\it {CP violation with
  beautiful baryons}}, }{}\href{http://dx.doi.org/10.1007/BF01589716}{Z.\
  Phys.\  {\bf C56} (1992) 129}\relax
\mciteBstWouldAddEndPuncttrue
\mciteSetBstMidEndSepPunct{\mcitedefaultmidpunct}
{\mcitedefaultendpunct}{\mcitedefaultseppunct}\relax
\EndOfBibitem
\bibitem{:1998upb}
Fayyazuddin, \ifthenelse{\boolean{articletitles}}{{\it {$\Lb \to \Lz +
  D^0(\overline{D}{}^0)$ decays and CP-violation}},
  }{}\href{http://dx.doi.org/10.1142/S0217732399000092}{Mod.\ Phys.\ Lett.\
  {\bf A14} (1999) 63}, \href{http://arxiv.org/abs/hep-ph/9806393}{{\tt
  arXiv:hep-ph/9806393}}\relax
\mciteBstWouldAddEndPuncttrue
\mciteSetBstMidEndSepPunct{\mcitedefaultmidpunct}
{\mcitedefaultendpunct}{\mcitedefaultseppunct}\relax
\EndOfBibitem
\bibitem{Giri:2001ju}
A.~K. Giri, R.~Mohanta, and M.~P. Khanna,
  \ifthenelse{\boolean{articletitles}}{{\it {Possibility of extracting the weak
  phase $\gamma$ from $\Lb \to \Lz D^0$ decays}},
  }{}\href{http://dx.doi.org/10.1103/PhysRevD.65.073029}{Phys.\ Rev.\  {\bf
  D65} (2002) 073029}, \href{http://arxiv.org/abs/hep-ph/0112220}{{\tt
  arXiv:hep-ph/0112220}}\relax
\mciteBstWouldAddEndPuncttrue
\mciteSetBstMidEndSepPunct{\mcitedefaultmidpunct}
{\mcitedefaultendpunct}{\mcitedefaultseppunct}\relax
\EndOfBibitem
\bibitem{Dunietz:1991yd}
I.~Dunietz, \ifthenelse{\boolean{articletitles}}{{\it {CP violation with
  self-tagging $B_d$ modes}},
  }{}\href{http://dx.doi.org/10.1016/0370-2693(91)91542-4}{Phys.\ Lett.\  {\bf
  B270} (1991) 75}\relax
\mciteBstWouldAddEndPuncttrue
\mciteSetBstMidEndSepPunct{\mcitedefaultmidpunct}
{\mcitedefaultendpunct}{\mcitedefaultseppunct}\relax
\EndOfBibitem
\bibitem{Gershon:2008pe}
T.~Gershon, \ifthenelse{\boolean{articletitles}}{{\it {On the measurement of
  the unitarity triangle angle $\gamma$ from $B^0 \to DK^{*0}$ decays}},
  }{}\href{http://dx.doi.org/10.1103/PhysRevD.79.051301}{Phys.\ Rev.\  {\bf
  D79} (2009) 051301}, \href{http://arxiv.org/abs/0810.2706}{{\tt
  arXiv:0810.2706}}\relax
\mciteBstWouldAddEndPuncttrue
\mciteSetBstMidEndSepPunct{\mcitedefaultmidpunct}
{\mcitedefaultendpunct}{\mcitedefaultseppunct}\relax
\EndOfBibitem
\bibitem{Gershon:2009qc}
T.~Gershon and M.~Williams, \ifthenelse{\boolean{articletitles}}{{\it
  {Prospects for the measurement of the unitarity triangle angle $\gamma$ from
  $B^0 \to DK^+\pim$ decays}},
  }{}\href{http://dx.doi.org/10.1103/PhysRevD.80.092002}{Phys.\ Rev.\  {\bf
  D80} (2009) 092002}, \href{http://arxiv.org/abs/0909.1495}{{\tt
  arXiv:0909.1495}}\relax
\mciteBstWouldAddEndPuncttrue
\mciteSetBstMidEndSepPunct{\mcitedefaultmidpunct}
{\mcitedefaultendpunct}{\mcitedefaultseppunct}\relax
\EndOfBibitem
\bibitem{Alves:2008zz}
LHCb collaboration, A.~A. Alves~Jr {\em et~al.},
  \ifthenelse{\boolean{articletitles}}{{\it {The \lhcb detector at the LHC}},
  }{}\href{http://dx.doi.org/10.1088/1748-0221/3/08/S08005}{JINST {\bf 3}
  (2008) S08005}\relax
\mciteBstWouldAddEndPuncttrue
\mciteSetBstMidEndSepPunct{\mcitedefaultmidpunct}
{\mcitedefaultendpunct}{\mcitedefaultseppunct}\relax
\EndOfBibitem
\bibitem{LHCb-PAPER-2012-012}
LHCb collaboration, R.~Aaij {\em et~al.},
  \ifthenelse{\boolean{articletitles}}{{\it {Observation of excited \Lb
  baryons}}, }{}\href{http://dx.doi.org/10.1103/PhysRevLett.109.172003}{Phys.\
  Rev.\ Lett.\  {\bf 109} (2012) 172003},
  \href{http://arxiv.org/abs/1205.3452}{{\tt arXiv:1205.3452}}\relax
\mciteBstWouldAddEndPuncttrue
\mciteSetBstMidEndSepPunct{\mcitedefaultmidpunct}
{\mcitedefaultendpunct}{\mcitedefaultseppunct}\relax
\EndOfBibitem
\bibitem{LHCb-CONF-2012-029}
{LHCb collaboration}, \ifthenelse{\boolean{articletitles}}{{\it {Measurement of
  the time-dependent \CP-violation parameters in $B^0_s \to D_s^\mp K^\pm$}},
  }{}
  \href{http://cdsweb.cern.ch/search?p={LHCb-CONF-2012-029}&f=reportnumber&act%
ion_search=Search&c=LHCb+Reports&c=LHCb+Conference+Proceedings&c=LHCb+Conferen%
ce+Contributions&c=LHCb+Notes&c=LHCb+Theses&c=LHCb+Papers}
  {{LHCb-CONF-2012-029}}\relax
\mciteBstWouldAddEndPuncttrue
\mciteSetBstMidEndSepPunct{\mcitedefaultmidpunct}
{\mcitedefaultendpunct}{\mcitedefaultseppunct}\relax
\EndOfBibitem
\bibitem{LHCb-DP-2012-003}
M.~Adinolfi {\em et~al.}, \ifthenelse{\boolean{articletitles}}{{\it
  {Performance of the \lhcb RICH detector at the LHC}},
  }{}\href{http://dx.doi.org/10.1140/epjc/s10052-013-2431-9}{Eur.\ Phys.\ J.\
  {\bf C73} (2013) 2431}, \href{http://arxiv.org/abs/1211.6759}{{\tt
  arXiv:1211.6759}}\relax
\mciteBstWouldAddEndPuncttrue
\mciteSetBstMidEndSepPunct{\mcitedefaultmidpunct}
{\mcitedefaultendpunct}{\mcitedefaultseppunct}\relax
\EndOfBibitem
\bibitem{LHCb-DP-2012-002}
A.~A. Alves~Jr {\em et~al.}, \ifthenelse{\boolean{articletitles}}{{\it
  {Performance of the LHCb muon system}},
  }{}\href{http://dx.doi.org/10.1088/1748-0221/8/02/P02022}{JINST {\bf 8}
  (2013) P02022}, \href{http://arxiv.org/abs/1211.1346}{{\tt
  arXiv:1211.1346}}\relax
\mciteBstWouldAddEndPuncttrue
\mciteSetBstMidEndSepPunct{\mcitedefaultmidpunct}
{\mcitedefaultendpunct}{\mcitedefaultseppunct}\relax
\EndOfBibitem
\bibitem{LHCb-DP-2012-004}
R.~Aaij {\em et~al.}, \ifthenelse{\boolean{articletitles}}{{\it {The \lhcb
  trigger and its performance in 2011}},
  }{}\href{http://dx.doi.org/10.1088/1748-0221/8/04/P04022}{JINST {\bf 8}
  (2013) P04022}, \href{http://arxiv.org/abs/1211.3055}{{\tt
  arXiv:1211.3055}}\relax
\mciteBstWouldAddEndPuncttrue
\mciteSetBstMidEndSepPunct{\mcitedefaultmidpunct}
{\mcitedefaultendpunct}{\mcitedefaultseppunct}\relax
\EndOfBibitem
\bibitem{BBDT}
V.~V. Gligorov and M.~Williams, \ifthenelse{\boolean{articletitles}}{{\it
  {Efficient, reliable and fast high-level triggering using a bonsai boosted
  decision tree}},
  }{}\href{http://dx.doi.org/10.1088/1748-0221/8/02/P02013}{JINST {\bf 8}
  (2013) P02013}, \href{http://arxiv.org/abs/1210.6861}{{\tt
  arXiv:1210.6861}}\relax
\mciteBstWouldAddEndPuncttrue
\mciteSetBstMidEndSepPunct{\mcitedefaultmidpunct}
{\mcitedefaultendpunct}{\mcitedefaultseppunct}\relax
\EndOfBibitem
\bibitem{Sjostrand:2006za}
T.~Sj\"{o}strand, S.~Mrenna, and P.~Skands,
  \ifthenelse{\boolean{articletitles}}{{\it {PYTHIA 6.4 physics and manual}},
  }{}\href{http://dx.doi.org/10.1088/1126-6708/2006/05/026}{JHEP {\bf 05}
  (2006) 026}, \href{http://arxiv.org/abs/hep-ph/0603175}{{\tt
  arXiv:hep-ph/0603175}}\relax
\mciteBstWouldAddEndPuncttrue
\mciteSetBstMidEndSepPunct{\mcitedefaultmidpunct}
{\mcitedefaultendpunct}{\mcitedefaultseppunct}\relax
\EndOfBibitem
\bibitem{LHCb-PROC-2010-056}
I.~Belyaev {\em et~al.}, \ifthenelse{\boolean{articletitles}}{{\it {Handling of
  the generation of primary events in \gauss, the \lhcb simulation framework}},
  }{}\href{http://dx.doi.org/10.1109/NSSMIC.2010.5873949}{Nuclear Science
  Symposium Conference Record (NSS/MIC) {\bf IEEE} (2010) 1155}\relax
\mciteBstWouldAddEndPuncttrue
\mciteSetBstMidEndSepPunct{\mcitedefaultmidpunct}
{\mcitedefaultendpunct}{\mcitedefaultseppunct}\relax
\EndOfBibitem
\bibitem{Lange:2001uf}
D.~J. Lange, \ifthenelse{\boolean{articletitles}}{{\it {The EvtGen particle
  decay simulation package}},
  }{}\href{http://dx.doi.org/10.1016/S0168-9002(01)00089-4}{Nucl.\ Instrum.\
  Meth.\  {\bf A462} (2001) 152}\relax
\mciteBstWouldAddEndPuncttrue
\mciteSetBstMidEndSepPunct{\mcitedefaultmidpunct}
{\mcitedefaultendpunct}{\mcitedefaultseppunct}\relax
\EndOfBibitem
\bibitem{Allison:2006ve}
Geant4 collaboration, J.~Allison {\em et~al.},
  \ifthenelse{\boolean{articletitles}}{{\it {Geant4 developments and
  applications}}, }{}\href{http://dx.doi.org/10.1109/TNS.2006.869826}{IEEE
  Trans.\ Nucl.\ Sci.\  {\bf 53} (2006) 270}\relax
\mciteBstWouldAddEndPuncttrue
\mciteSetBstMidEndSepPunct{\mcitedefaultmidpunct}
{\mcitedefaultendpunct}{\mcitedefaultseppunct}\relax
\EndOfBibitem
\bibitem{Agostinelli:2002hh}
Geant4 collaboration, S.~Agostinelli {\em et~al.},
  \ifthenelse{\boolean{articletitles}}{{\it {Geant4: a simulation toolkit}},
  }{}\href{http://dx.doi.org/10.1016/S0168-9002(03)01368-8}{Nucl.\ Instrum.\
  Meth.\  {\bf A506} (2003) 250}\relax
\mciteBstWouldAddEndPuncttrue
\mciteSetBstMidEndSepPunct{\mcitedefaultmidpunct}
{\mcitedefaultendpunct}{\mcitedefaultseppunct}\relax
\EndOfBibitem
\bibitem{LHCb-PROC-2011-006}
M.~Clemencic {\em et~al.}, \ifthenelse{\boolean{articletitles}}{{\it {The \lhcb
  simulation application, \gauss: design, evolution and experience}},
  }{}\href{http://dx.doi.org/10.1088/1742-6596/331/3/032023}{{J.\ Phys.\ Conf.\
  Ser.\ } {\bf 331} (2011) 032023}\relax
\mciteBstWouldAddEndPuncttrue
\mciteSetBstMidEndSepPunct{\mcitedefaultmidpunct}
{\mcitedefaultendpunct}{\mcitedefaultseppunct}\relax
\EndOfBibitem
\bibitem{Hulsbergen:2005pu}
W.~D. Hulsbergen, \ifthenelse{\boolean{articletitles}}{{\it {Decay chain
  fitting with a Kalman filter}},
  }{}\href{http://dx.doi.org/10.1016/j.nima.2005.06.078}{Nucl.\ Instrum.\
  Meth.\  {\bf A552} (2005) 566},
  \href{http://arxiv.org/abs/physics/0503191}{{\tt
  arXiv:physics/0503191}}\relax
\mciteBstWouldAddEndPuncttrue
\mciteSetBstMidEndSepPunct{\mcitedefaultmidpunct}
{\mcitedefaultendpunct}{\mcitedefaultseppunct}\relax
\EndOfBibitem
\bibitem{LHCb-PAPER-2012-048}
LHCb collaboration, R.~Aaij {\em et~al.},
  \ifthenelse{\boolean{articletitles}}{{\it {Measurements of the $\Lambda_b^0$,
  $\Xi_b^-$ and $\Omega_b^-$ baryon masses}},
  }{}\href{http://dx.doi.org/10.1103/PhysRevLett.110.182001}{Phys.\ Rev.\
  Lett.\  {\bf 110} (2013) 182001}, \href{http://arxiv.org/abs/1302.1072}{{\tt
  arXiv:1302.1072}}\relax
\mciteBstWouldAddEndPuncttrue
\mciteSetBstMidEndSepPunct{\mcitedefaultmidpunct}
{\mcitedefaultendpunct}{\mcitedefaultseppunct}\relax
\EndOfBibitem
\bibitem{Breiman}
L.~Breiman, J.~H. Friedman, R.~A. Olshen, and C.~J. Stone, {\em Classification
  and regression trees}, Wadsworth international group, Belmont, California,
  USA, 1984\relax
\mciteBstWouldAddEndPuncttrue
\mciteSetBstMidEndSepPunct{\mcitedefaultmidpunct}
{\mcitedefaultendpunct}{\mcitedefaultseppunct}\relax
\EndOfBibitem
\bibitem{Skwarnicki:1986xj}
T.~Skwarnicki, {\em {A study of the radiative cascade transitions between the
  Upsilon-prime and Upsilon resonances}}, PhD thesis, Institute of Nuclear
  Physics, Krakow, 1986,
  {\href{http://inspirehep.net/record/230779/files/230779.pdf}{DESY-F31-86-02}%
}\relax
\mciteBstWouldAddEndPuncttrue
\mciteSetBstMidEndSepPunct{\mcitedefaultmidpunct}
{\mcitedefaultendpunct}{\mcitedefaultseppunct}\relax
\EndOfBibitem
\bibitem{LHCb-PAPER-2012-018}
LHCb collaboration, R.~Aaij {\em et~al.},
  \ifthenelse{\boolean{articletitles}}{{\it {Observation of $B^0 \to
  \overline{D}^0 K^+ K^-$ and evidence for $B^0_s \to \overline{D}^0 K^+
  K^-$}}, }{}\href{http://dx.doi.org/10.1103/PhysRevLett.109.131801}{Phys.\
  Rev.\ Lett.\  {\bf 109} (2012) 131801},
  \href{http://arxiv.org/abs/1207.5991}{{\tt arXiv:1207.5991}}\relax
\mciteBstWouldAddEndPuncttrue
\mciteSetBstMidEndSepPunct{\mcitedefaultmidpunct}
{\mcitedefaultendpunct}{\mcitedefaultseppunct}\relax
\EndOfBibitem
\bibitem{LHCb-PAPER-2013-022}
LHCb collaboration, R.~Aaij {\em et~al.},
  \ifthenelse{\boolean{articletitles}}{{\it {Measurements of the branching
  fractions of the decays $B^0_s \rightarrow \overline{D}^0 K^- \pi^+$ and $B^0
  \rightarrow \overline{D}^0 K^+ \pi^-$}},
  }{}\href{http://dx.doi.org/10.1103/PhysRevD.87.112009}{Phys.\ Rev.\  {\bf
  D87} (2013) 112009}, \href{http://arxiv.org/abs/1304.6317}{{\tt
  arXiv:1304.6317}}\relax
\mciteBstWouldAddEndPuncttrue
\mciteSetBstMidEndSepPunct{\mcitedefaultmidpunct}
{\mcitedefaultendpunct}{\mcitedefaultseppunct}\relax
\EndOfBibitem
\bibitem{LHCb-PAPER-2012-037}
LHCb collaboration, R.~Aaij {\em et~al.},
  \ifthenelse{\boolean{articletitles}}{{\it {Measurement of the fragmentation
  fraction ratio $f_s/f_d$ and its dependence on $B$ meson kinematics}},
  }{}\href{http://dx.doi.org/10.1007/JHEP04(2013)001}{JHEP {\bf 04} (2013) 1},
  \href{http://arxiv.org/abs/1301.5286}{{\tt arXiv:1301.5286}}\relax
\mciteBstWouldAddEndPuncttrue
\mciteSetBstMidEndSepPunct{\mcitedefaultmidpunct}
{\mcitedefaultendpunct}{\mcitedefaultseppunct}\relax
\EndOfBibitem
\bibitem{Pivk:2004ty}
M.~Pivk and F.~R. Le~Diberder, \ifthenelse{\boolean{articletitles}}{{\it
  {sPlot: a statistical tool to unfold data distributions}},
  }{}\href{http://dx.doi.org/10.1016/j.nima.2005.08.106}{Nucl.\ Instrum.\
  Meth.\  {\bf A555} (2005) 356},
  \href{http://arxiv.org/abs/physics/0402083}{{\tt
  arXiv:physics/0402083}}\relax
\mciteBstWouldAddEndPuncttrue
\mciteSetBstMidEndSepPunct{\mcitedefaultmidpunct}
{\mcitedefaultendpunct}{\mcitedefaultseppunct}\relax
\EndOfBibitem
\bibitem{Aitala:1999uq}
E791 collaboration, E.~Aitala {\em et~al.},
  \ifthenelse{\boolean{articletitles}}{{\it {Multidimensional resonance
  analysis of $\Lc \to p K^- \pip$}},
  }{}\href{http://dx.doi.org/10.1016/S0370-2693(99)01397-0}{Phys.\ Lett.\  {\bf
  B471} (2000) 449}, \href{http://arxiv.org/abs/hep-ex/9912003}{{\tt
  arXiv:hep-ex/9912003}}\relax
\mciteBstWouldAddEndPuncttrue
\mciteSetBstMidEndSepPunct{\mcitedefaultmidpunct}
{\mcitedefaultendpunct}{\mcitedefaultseppunct}\relax
\EndOfBibitem
\bibitem{LHCb-PAPER-2012-057}
LHCb collaboration, R.~Aaij {\em et~al.},
  \ifthenelse{\boolean{articletitles}}{{\it {Measurements of the $\Lb \to \Lz
  J/\psi$ decay amplitudes and the $\Lb$ baryon production polarisation in $pp$
  collisions at $\sqrt{s} = 7\tev$}},
  }{}\href{http://dx.doi.org/10.1016/j.physletb.2013.05.041}{Phys.\ Lett.\
  {\bf B724} (2013) 27}, \href{http://arxiv.org/abs/1302.5578}{{\tt
  arXiv:1302.5578}}\relax
\mciteBstWouldAddEndPuncttrue
\mciteSetBstMidEndSepPunct{\mcitedefaultmidpunct}
{\mcitedefaultendpunct}{\mcitedefaultseppunct}\relax
\EndOfBibitem
\bibitem{LHCb-PAPER-2013-032}
LHCb collaboration, R.~Aaij {\em et~al.},
  \ifthenelse{\boolean{articletitles}}{{\it {Precision measurement of the $\Lb$
  baryon lifetime}},
  }{}\href{http://dx.doi.org/10.1103/PhysRevLett.111.102003}{Phys.\ Rev.\
  Lett.\  {\bf 111} (2013) 102003}, \href{http://arxiv.org/abs/1307.2476}{{\tt
  arXiv:1307.2476}}\relax
\mciteBstWouldAddEndPuncttrue
\mciteSetBstMidEndSepPunct{\mcitedefaultmidpunct}
{\mcitedefaultendpunct}{\mcitedefaultseppunct}\relax
\EndOfBibitem
\end{mcitethebibliography}

\end{document}